\documentclass[pdflatex,sn-mathphys-num]{sn-jnl}%

\usepackage{graphicx}%
\usepackage{multirow}%
\usepackage{amsmath,amssymb,amsfonts}%
\usepackage{amsthm}%
\usepackage{mathrsfs}%
\usepackage[title]{appendix}%
\usepackage{xcolor}%
\usepackage{textcomp}%
\usepackage{manyfoot}%
\usepackage{booktabs}%
\usepackage[normalem]{ulem}
\usepackage{algorithm}%
\usepackage{algorithmicx}%
\usepackage{algpseudocode}%
\usepackage{listings}%
\usepackage{amsmath}

\newcommand{\fig}{\text{Fig.~}}
\newcommand{\eq}{\text{Eq.~}}
\DeclareMathOperator*{\argmin}{arg\,min}
\DeclareMathOperator*{\argmax}{arg\,max}

\unnumbered

\begin{document}

\title{Cognitive behavior emerging in mechanical systems through reservoir encoding and active inference}

\author*[1]{\fnm{Matteo} \sur{Torzoni}}\email{matteo.torzoni@polimi.it}

\author[2]{\fnm{Domenico} \sur{Maisto}}\email{domenico.maisto@istc.cnr.it}

\author[3]{\fnm{Andrea} \sur{Manzoni}}\email{andrea1.manzoni@polimi.it}

\author[2]{\fnm{Francesco} \sur{Donnarumma}}\email{francesco.donnarumma@istc.cnr.it}

\author[2]{\fnm{Giovanni} \sur{Pezzulo}}\email{giovanni.pezzulo@istc.cnr.it}

\author[1]{\fnm{Alberto} \sur{Corigliano}}\email{alberto.corigliano@polimi.it}

\affil*[1]{\orgdiv{Department of Civil and Environmental Engineering}, \orgname{Politecnico di Milano}, \orgaddress{\city{Milano}, \country{Italy}}}

\affil[2]{\orgdiv{Institute of Cognitive Sciences and Technologies}, \orgname{National Research Council}, \orgaddress{\city{Roma}, \country{Italy}}}

\affil[3]{\orgdiv{MOX-Department of
Mathematics}, \orgname{Politecnico di Milano}, \orgaddress{\city{Milano}, \country{Italy}}}

\abstract{Mechanical systems are traditionally designed as passive assets, where sensing, analysis, and control are treated as decoupled and externally prescribed processes. Reliance on predefined mechanistic models and supervised learning with offline optimization further limits their adaptability in changing environments. Drawing inspiration from biological cognition, we introduce a computational framework enabling mechanical systems to learn, adapt, and make decisions through interactions with partially observable environments. Perception, action, and learning are integrated through active inference within a perception-action loop. Sensor data are assimilated through a neural reservoir to support continual belief updating, while actions actively probe and influence the physical system, becoming an integral part of the inference process. Through self-directed interaction, adaptive behavior emerges from the continual refinement of internal representations to explain varying observation patterns while pursuing mechanical objectives. Across simulated case studies involving multiple mechanical systems, static and dynamic sensing, and objectives ranging from deformation maximization to vibration mitigation, we demonstrate the potential of the proposed framework as enabling paradigm for autonomous mechanical systems.}

\keywords{Active inference, Reservoir computing, Cognitive mechanical systems}

\maketitle

\section{Introduction}\label{sec1}
Structural optimization, monitoring, and control are exemplary engineering applications where sensing, analysis, and actuation are typically treated as decoupled, externally prescribed processes. This perspective originates from the traditional view of mechanical systems as passive assets operating within predefined functional regimes. Here we propose a fundamentally different framework for intelligent automation in mechanical systems~\cite{san2026evolution}. We draw inspiration from recent efforts in theoretical biology and neuroscience to understand the emergence of embodied intelligence and adaptive behavior across scales--not only in advanced animals, but also in simpler forms of life, individual biological cells, microbial communities, organoids, collective systems, and, more broadly, in artificial systems, algorithms, and smart materials~\cite{pezzulo2015active,tiwary2025if,pfeifer2006body,levin2022technological,pezzulo2026bootstrapping,marom2025frontiers}. In keeping with this perspective, here we propose a computational framework to endow materials and structures with cognitive-inspired capabilities for learning and environmental adaptation.

The proposed framework enables modeling and developing continuous closed-loop interactions between smart mechanical systems and their surroundings. Information flows from the physical system to its coupled digital representation through the acquisition and elaboration of sensor data, and back to the physical domain through actuation and observation. Learning and adaptation emerge from the continual refinement of internal representations required to explain incoming observations under changing operational conditions. This ongoing regulation gives rise to both reactive and anticipatory behaviors, enabling the system to pursue mechanical objectives without requiring mechanistic models of the underlying physical phenomena.

Recent advances in smart materials~\cite{de2026multimodal,cummer2016controlling,PontiPhysRevApplied,wu2024wave}, sensing technology~\cite{ML_perspective,corigliano2018mechanics,lynch2006review}, data-driven modeling ~\cite{ficili2025sensors,Alu_2025,herrmann2024deep,karniadakis2021physics,LIU2022109276}, and control~\cite{naughton2026neural,hu2025teaching,Brunton,pfeifer2007self,andriotis2019managing,liu2025embodied} are expanding the ability of mechanical systems to respond to changing operational conditions. For instance, in structural monitoring, model-based approaches typically address the ill-posed nature of data assimilation through Bayesian model updating and uncertainty quantification~\cite{Muto,AM_Green}, whereas data-driven approaches rely on a pattern-recognition paradigm involving operational evaluation, data acquisition, feature extraction, and statistical modeling~\cite{Farrar01}. In parallel, digital twins are emerging as platforms that integrate observational data with computational models to enable predictive and diagnostic capabilities beyond those achievable through either component alone~\cite{Foundational,pgm_wilcox_dt,henneking2026goal,Torzoni_DT}. Nevertheless, most current implementations are still rooted in reactive operation and supervised learning, relying on predefined models or offline datasets. As a result, adaptation to evolving environments remains limited, while sensing, data assimilation, learning, and control are typically treated as separate processes~\cite{MT_AIF}.

Adaptive behavior in cognitive systems emerges from continuous interactions with the environment, where perception, action, and learning are intertwined aspects of a single dynamical process~\cite{Clark2016,beer2000dynamical}. In this context, cognitive behavior does not necessarily imply human-like intelligence, but rather the capacity for adaptive agency through active engagement with the environment to sense, infer hidden causes relevant for prediction and decision-making, and select actions that achieve preferred outcomes. This perspective has been formalized in computational neuroscience through predictive coding-inspired frameworks such as active inference (AIF)~\cite{parr2022active}. Within AIF, agents continuously update internal beliefs to explain sensory observations under uncertainty, while simultaneously selecting actions both to gather informative evidence and steer the environment toward preferred outcomes. This perception--action loop is formulated through the minimization of variational free energy, a probabilistic objective that quantifies prediction uncertainty together with the consistency between internal beliefs, observations, and behavioral preferences~\cite{friston2010free,buckley2017free}.

Beyond neuroscience~\cite{Friston02102015,han2024synergizing,fitzgerald2015dopamine,isomura2018vitro,van2024hierarchical,pezzulo2024neural}, AIF has been applied to decision-making~\cite{fountas2020deep,mazzaglia2021contrastive,hashash2026active}, robotics~\cite{taniguchi2023world,vijayaraghavan2025development}, and collective behavior~\cite{maisto2023interactive,heins2024collective}. It also shares conceptual similarities with the ``dynamic data-driven application systems'' paradigm~\cite{darema2004dynamic}. However, unlike reactive approaches, AIF incorporates active information seeking as a mechanism for improving future performance~\cite{friston2017graphical}. Recent studies have begun to explore these ideas in engineering contexts~\cite{MT_AIF,10502828,de2026active,Deep_kin}, highlighting a transition toward cognitive-inspired frameworks. 

\begin{figure}[!t]
\centering\includegraphics[width=1\linewidth]{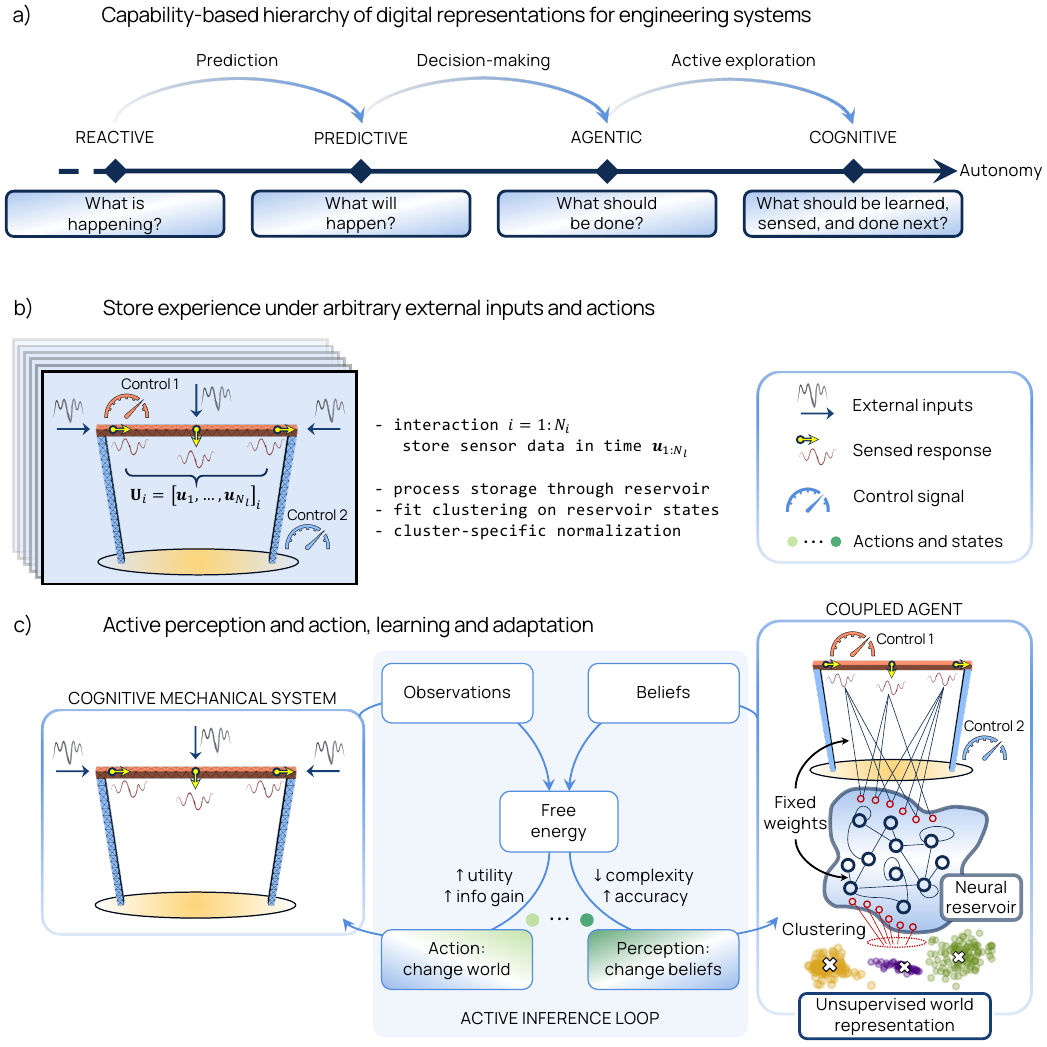}
\caption{\textbf{Graphical abstraction of a cognitive mechanical system and its coupled active inference agent.} \textbf{a} Evolution of digital representations through reactive, predictive, agentic, and cognitive frameworks. \textbf{b} Experience generation. The mechanical system is exposed to arbitrary external inputs and random exploratory actions to acquire an initial experience base. \textbf{c} Active perception and action, learning and adaptation. The mechanical system, equipped with sensors and actuators, is modeled as a partially observable dynamical environment. Incoming measurements are encoded through a fixed neural reservoir into latent representations, subsequently clustered to define discrete observations supporting contextual inference. The active inference agent updates beliefs about hidden states while selecting actions that simultaneously improve future observations and achieve mechanical objectives by minimizing \textit{expected free energy}. Initialized with an uninformative observation model, the agent progressively refines its internal representations through self-supervised interaction within the perception--action loop, allowing hidden states to acquire semantic meaning without predefined labels, mechanistic models, or rigid separation between data collection, learning, and deployment.\\}
\label{fig:fig1}
\end{figure}

Figure~\ref{fig:fig1}a summarizes this evolution through a capability-based hierarchy of digital representations, from reactive to cognitive systems. Reactive representations primarily support state estimation and monitoring through continuous data assimilation. Predictive representations extend these capabilities by forecasting future behavior through the integration of observational data and computational models. Agentic representations incorporate autonomous decision-making, enabling the comparison of alternative actions and their consequences. The proposed cognitive framework further integrates continual adaptation through self-supervised learning driven by active information gathering. This closes the perception--action loop, allowing internal representations to be continuously refined in pursuit of operational objectives, rather than being prescribed from predefined mechanistic models.

In keeping with this perspective, here we model (simulated) cognitive mechanical systems as partially observable dynamical environments~\cite{planning_acting}, endowed with a coupled AIF agent whose internal representations continuously evolve through self-directed interaction. In doing so, the proposed framework embraces and overcomes constraints that are often simplified or treated separately in existing methodologies, including the absence of supervision, continual adaptation, and limited computational resources. An initial experience base is acquired through blind interaction with the physical system under arbitrary external inputs and random actions (\fig\ref{fig:fig1}b). During operation, structural signals are transformed into reservoir-based representations that capture contextual information under partial observability (\fig\ref{fig:fig1}c). Reservoir computing is particularly attractive in this context because it processes temporal signals through fixed nonlinear dynamics, avoiding the costly training required by neural networks while maintaining adaptability under evolving conditions~\cite{Jaeger,maass2002real,nakajima2021reservoir,yan2024emerging}. Reservoir representations are then clustered into a finite observation space, converting sensor measurements into observations suitable for probabilistic inference about hidden system states under discrete state-space models. Free-energy minimization closes the perception--action loop by selecting actions that simultaneously improve future perception and achieve mechanical objectives. 

Unlike conventional machine learning workflows, no separation is imposed between data collection, learning, and deployment. The agent is initialized without prior knowledge of the causal relationships between hidden states and observations, reflecting no prior knowledge about the underlying structural behavior. Neither predefined labels, nor physics-based models of the underlying phenomena, are required. Instead, internal representations are progressively learned through self-supervised interaction within the perception--action loop, while hidden states acquire semantic meaning through their inferred relationships with observations, actions, and outcomes. Adaptation to previously unseen operational regimes is enabled through state expansion~\cite{smith2020active}, allowing novel scenarios to be incorporated without disrupting previously acquired knowledge. Online learning is further complemented by offline Bayesian model reduction (BMR)~\cite{friston2017active}, which subsequently compresses the learned model by balancing explanatory accuracy against model complexity.

Following these premises, we evaluate the proposed framework on two complementary numerical demonstration platforms, each coupling a simulated mechanical system to an AIF agent. The first combines static displacement or strain sensing with the control of an operational variable under hidden boundary conditions, whereas the second combines dynamic acceleration sensing with the control of multiple mechanical-property variables under changing loads and competing performance and resource objectives. These settings are relevant, respectively, to tunable structural systems operating under changing boundary conditions and resource-aware vibration control under variable dynamic loads. Together, these case studies assess whether the same perception-action-learning framework can accommodate different observation structures, action spaces, and mechanical objectives using a common set of core hyperparameters. Beyond numerical validation, the proposed framework suggests unprecedented possibilities for engineering systems. Indeed, it transcends simulation and offers the potential to serve as a blueprint for autonomous mechanical systems.

\section{Results}\label{sec2}
This section presents a series of simulation-based case studies illustrating how adaptive behavior can emerge in cognitive mechanical systems endowed with a coupled AIF agent. In all case studies, the physical system, or \textit{generative process}, is only partially observable by the agent through sparse sensor measurements. To demonstrate the generality of the proposed framework, we consider both static and dynamic sensing, as well as distinct mechanical objectives ranging from deformation maximization to resource-constrained vibration mitigation.

\begin{figure}[!t]
\centering\includegraphics[width=0.94\linewidth]{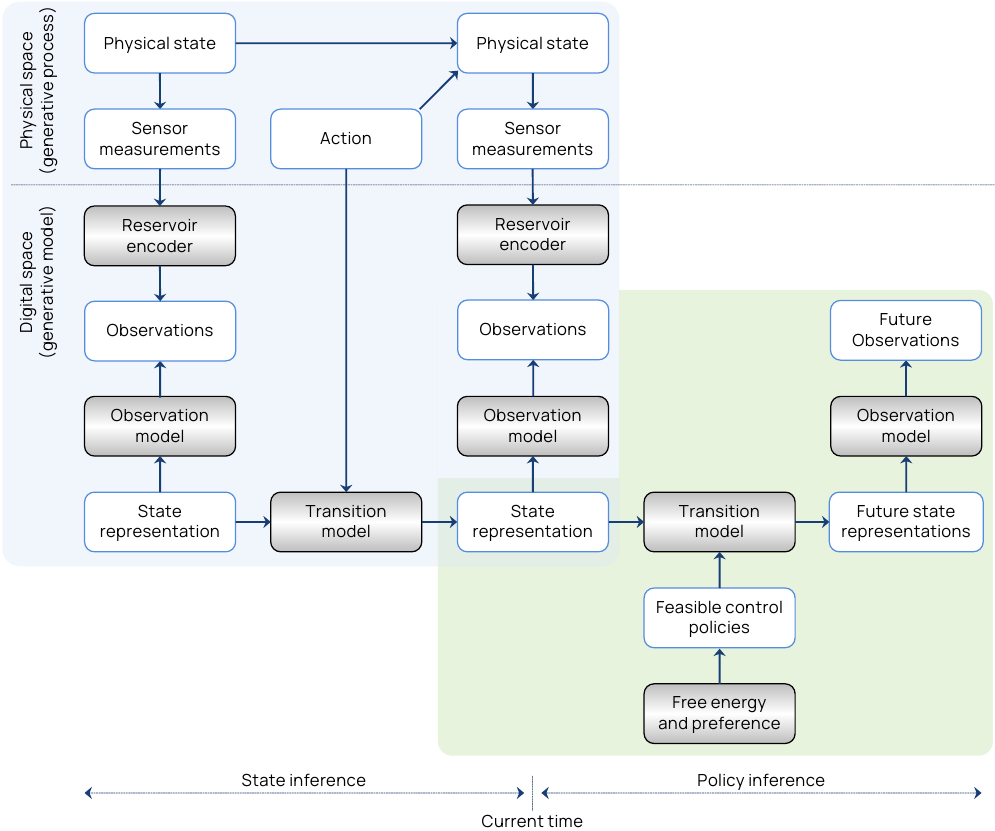}
\caption{\textbf{Graph-like schematic of the agent--environment interaction.} The horizontal line separates the external physical process generating sensor data (the \textit{generative process}) from the agent's internal probabilistic representation (the \textit{generative model}). White boxes denote random variables, whereas gray boxes denote parametrized components of the generative model. Directed edges indicate conditional dependencies. Along the upper pathway, the physical state evolves under the influence of external inputs and actions, generating noisy sensor measurements according to the unknown generative process. These measurements are transformed into a discrete reservoir encoding, which is subsequently interpreted as observations by the active inference agent. The generative model describes how hidden states generate these observations through the observation model and evolve under candidate actions through the transition model. Blue and green background shadings identify the two complementary inference processes: state estimation and policy selection. Variational message passing continuously updates beliefs over hidden states, while policy inference evaluates candidate control actions by minimizing expected free energy.}
\label{fig:fig2}
\end{figure}

In all case studies, the agent is equipped with a discrete AIF \textit{generative model} (schematically illustrated in \fig\ref{fig:fig2}; see Section ``Active inference'' of the Methods for details). The generative model specifies how hidden states generate observations and evolve under the influence of actions. This should not be confused with a numerical model of the mechanical system: it is a probabilistic internal representation learned by the agent to capture the relationships between latent states, observations, and actions, rather than a mechanistic description of the underlying physics. During interaction, the agent continuously updates its beliefs about these latent states, while selecting actions that simultaneously improve future observations and achieve the desired mechanical objectives to minimize expected free energy.

The agent receives two complementary and interdependent observation modalities. The first consists of clustered reservoir representations, obtained by processing raw structural measurements through the neural reservoir and clustering the resulting embeddings. This modality captures latent spatiotemporal patterns to support the inference of the underlying operating regime (full details of the neural reservoir architecture are provided in Section ``Neural reservoir'' of the Methods). The second modality is derived from a norm of the raw sensor measurements and 
depends directly on the first, as signal norms are normalized and discretized conditionally on the inferred reservoir cluster. Rather than using fixed thresholds, this conditional structure allows preferences associated with the mechanical behavior to be evaluated relative to the specific inferred context. In this way, the reservoir encoding provides the structural context, while the signal norms acquire physical meaning only within that framework, mirroring the context-dependent nature of biological perception. This choice accommodates the fact that different operating regimes may produce sensor measurements with varying characteristics and magnitudes, thereby allowing the framework to contextualize the value of observations across different behavioral regimes.

Before entering active operation, the system undergoes a pre-learning phase driven by arbitrary external inputs and random actions to generate an initial experience base. This information is used to construct the reservoir clusters and estimate cluster-specific normalization statistics. The agent is then initialized with an uninformative observation model, reflecting no prior knowledge of how hidden states generate observations. As interactions accumulate, the observation model is progressively refined through self-supervised learning within the perception--action loop, while hidden states acquire semantic meaning through their inferred relationships with observations, actions, and outcomes.

The virtual environments emulating the physical systems and supporting interaction with the AIF agents have been implemented using the \texttt{Python} computing platform \texttt{DOLFINx}~\cite{dolfin}. Discrete Markovian AIF agents have been implemented using the \texttt{Python} library \texttt{pymdp}~\cite{pymdp}. State and policy inference are performed through variational message passing on a factorized probabilistic graphical model under mean-field approximation. Online learning of the observation models is achieved through Bayesian updates of Dirichlet concentration parameters (see Section ``Active inference'' of the Methods for implementation details). 

The resulting framework maintains a low computational cost, allowing all test cases considered in this work to run on laptop-class hardware. The framework also requires minimal hyperparameter tuning. Indeed, all the presented test cases rely on the following common set of settings: an exploratory phase of $1000$ interactions; uniform observation-model priors perturbed by zero-mean Gaussian noise with variance $5\cdot10^{-4}$; learning rate $0.6$; a reservoir with a size of $400$ neurons, spectral radius $0.9$, and connection density $0.1$. 

\subsection{Variable supports beam under movable load (Pinball beam)}
The first case study deals with positioning a mass along a beam to maximize its deformation under support conditions that change stochastically during operation. This setup reflects a broad class of engineering applications in which structures with tunable components must adapt to unknown or evolving operating conditions.

\begin{figure}[t!]
\centering\includegraphics[width=1\linewidth]{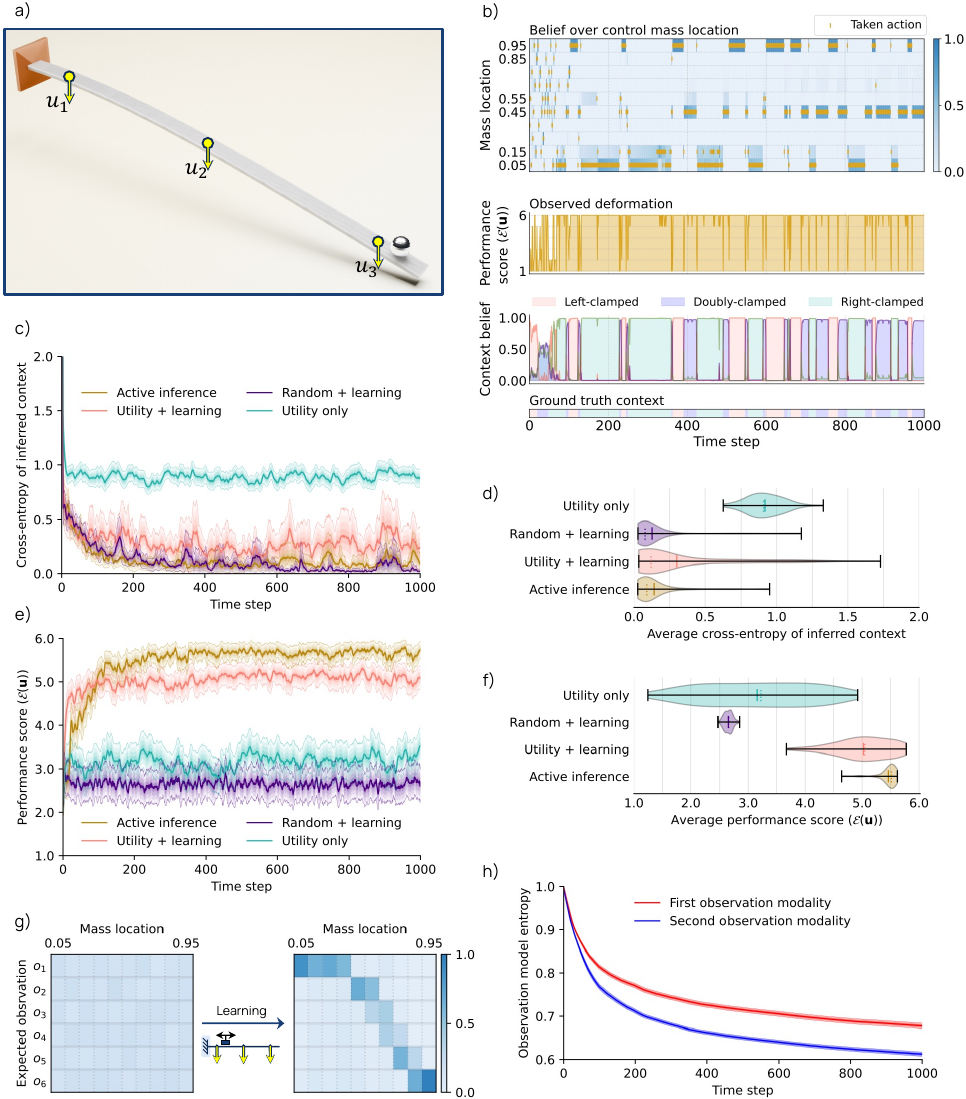}
\caption{\textbf{Case study 1: Pinball beam for the displacement-maximization objective.} \textbf{a} Schematic of the Pinball beam, showing the movable mass and three sensors providing sparse, noisy measurements ($u_1,u_2,u_3$) of the structural response. The support regime switches stochastically among left-clamped, doubly-clamped, and right-clamped configurations. \textbf{b} Representative evolution of the perception--action loop. From bottom to top: ground-truth support regime; inferred beliefs over the latent support context; observed deformation level defining the mechanical objective; and posterior beliefs over the movable-mass position, where color intensity denotes the probability assigned to each location and the gold marker indicates the selected action. A video showing the evolution of the perception--action loop synchronized with a rendering of the cognitive mechanical system is available in Movie~1 of the Supplementary Materials.}
\label{fig:beam_1}
\end{figure}

The system consists of an aluminum beam, shown in \fig\ref{fig:beam_1}a, of length $1~\mathrm{m}$, width $0.04~\mathrm{m}$, and thickness $0.01~\mathrm{m}$, with Young's modulus $70~\mathrm{GPa}$, Poisson's ratio $0.33$, and density $2700~\mathrm{kg/m^3}$. A distributed load equivalent to a mass of $1~\mathrm{kg}$, represented in \fig\ref{fig:beam_1}a as a ball, is applied over a region of dimensions $(0.1\times0.04)~\mathrm{m}^2$ on the beam top surface. The coupled agent controls the mass position over discrete locations uniformly distributed along the beam span, namely $\lbrace0.05~\mathrm{m},0.15~\mathrm{m},\ldots,0.95~\mathrm{m}\rbrace$, which define the action space.

\setcounter{figure}{2}    
\begin{figure}[t!]
\caption{\textbf{Caption continued.}  \textbf{c} Evolution of the contextual cross-entropy between the inferred beliefs and the ground-truth support regime for the proposed active inference framework and three reference strategies (utility + learning, random + learning, and utility only). Solid lines denote averages over $100$ independent runs with different random seeds, and shaded regions indicate $95\%$ confidence intervals. \textbf{d} Violin plots of the average contextual cross-entropy for the same $100$ runs. \textbf{e} Evolution of the mechanical objective, expressed as the mean squared sensor amplitude $\mathcal{E}(\mathbf{u})=\lVert\mathbf{u}\rVert_2^2/N_s$. \textbf{f} Violin plots of the average mechanical performance. \textbf{g} One slice of the initial and the learned observation model associated with the signal-norm observation modality, illustrating the emergence of meaningful associations between context slots, mass positions, and the structural response. \textbf{h} Evolution of the normalized entropy of the learned observation model for the reservoir-cluster (first) and signal-norm (second) observation modalities, demonstrating the progressive refinement of observation likelihoods from initially uninformative priors.}
\end{figure}

Support conditions switch stochastically among the left-clamped, doubly-clamped, and right-clamped configurations. At each physical--digital interaction, there is a probability of $1/30$ of switching from the current configuration to any of the two inactive configurations, while the remaining probability corresponds to retaining the current support condition. The (virtual) generative process yielding sensor data is modeled as a linear-elastic continuum under small-strain assumptions, discretized using linear hexahedron finite elements, and results in $1260$ degrees of freedom. The beam is observable only through $N_s=3$ noisy sensors positioned along its lower surface, providing sparse measurements $\mathbf{u}\in\mathbb{R}^{N_s}$. To emulate realistic sensing conditions, measurements are corrupted by additive zero-mean Gaussian noise with standard deviation set to $5\%$ of the signal magnitude.

Sensor measurements are processed through the neural reservoir and clustered to generate the first observation modality, facilitating inference of the underlying support context. The second modality consists of the mean squared sensor amplitude \mbox{$\mathcal{E}(\mathbf{u})=\lVert\mathbf{u}\rVert_2^2/N_s$}, enabling preferences to be expressed over the mechanical behavior.

The agent maintains two hidden-state factors describing the mass position and the hidden support context (see Section~A of the Supplementary Materials for details on the assigned initial prior over hidden states). The number of context slots can be specified independently of the actual number of physical support conditions, allowing the framework to operate without prior knowledge of the true environmental structure.

The observation model is initialized from uninformative Dirichlet priors and progressively refined through Bayesian learning from the perception--action loop. The transition dynamics is deterministic for the controllable mass position, whereas the latent support context evolves according to a simple persistence model that assigns a $90\%$ probability of remaining in the current state and distributes the remaining $10\%$ uniformly across the alternative context slots. Preferences are specified exclusively over the signal-norm observation modality to promote beam deformation (please see Section~A of the Supplementary Materials for implementation details and numerical values), while no preference is assigned over the reservoir-cluster modality.

Figure~\ref{fig:beam_1}b demonstrates the emergence of adaptive behavior for the displacement-maximization objective (please see also Movie~1 of the Supplementary Materials for a video showing the evolution of the perception--action loop synchronized with a rendering of the cognitive mechanical system). During the early stages of interaction, the agent actively explores the available actions to reduce uncertainty about its generative model. Through this process, the observation model progressively learns the relationships among latent context slots, mass positions, and the resulting structural response (\fig\ref{fig:beam_1}g,~h). Despite the continual switching of boundary conditions, beliefs over the hidden support regime become increasingly consistent with the true operating condition. Eventually, the agent spontaneously transitions from information-seeking to goal-directed behavior, progressively selecting mass positions that maximize the deformation of the beam.

A comparison with alternative agents further highlights the role of epistemic behavior. An agent combining utility maximization with online learning, but lacking active information seeking, may initially achieve comparable or even better performance owing to the simplicity of the environment and favorable random exploration (\fig\ref{fig:beam_1}e). However, lacking a mechanism to actively seek informative observations, its learning ultimately converges to a less effective control strategy (\fig\ref{fig:beam_1}e,~f). By contrast, the AIF agent autonomously switches from curiosity-driven information gathering to utility maximization as uncertainty over the observation model decreases. This adaptive transition enables superior long-term performance, yielding an average increase of about 15\% by the end of the simulation, while maintaining accurate contextual awareness (\fig\ref{fig:beam_1}c,~d). Note that this ablation study is intended to compare the components of the AIF framework only, isolating their individual contributions and demonstrating their complementary roles (please see Section~B of the Supplementary Materials for details on the evaluation metrics computation).

\begin{figure}[t!]
\centering\includegraphics[width=1\linewidth]{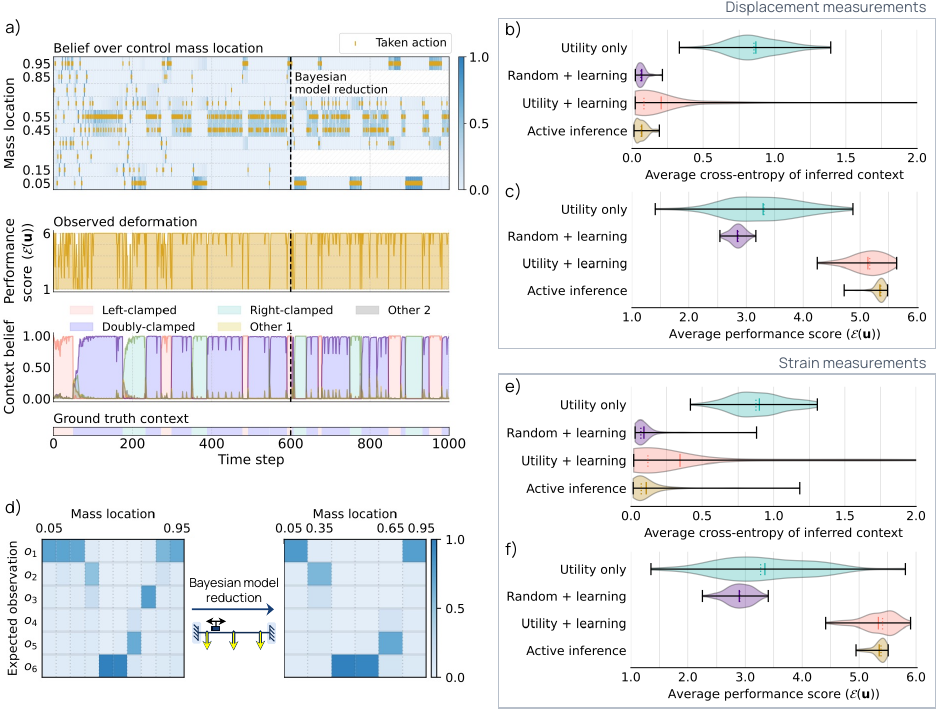}
\caption{\textbf{Case study 1: Overparametrized context representation and Bayesian model reduction.} \textbf{a} Representative evolution of the perception--action loop when the generative model is initialized with five latent context slots, despite the physical system exhibits three support regimes. From bottom to top: ground-truth support regime; inferred beliefs over the latent support context; observed deformation level defining the mechanical objective; and posterior beliefs over the movable-mass position, where color intensity denotes the probability assigned to each location and the gold marker indicates the selected action. At time step 600, Bayesian model reduction prunes redundant mass-location hypotheses while preserving the learned observation mappings. \textbf{b} Violin plots of the average contextual cross-entropy over $100$ independent runs with different random seeds for the proposed active inference framework and three reference strategies (utility + learning, random + learning, and utility only) using displacement measurements, demonstrating that accurate contextual representations emerge despite the overparametrized latent state space. \textbf{c} Violin plots of the average mechanical performance for the same $100$ runs. \textbf{d} Observation model associated with the signal-norm observation modality immediately before and after Bayesian model reduction, illustrating the removal of redundant mass-location hypotheses while preserving the learned mappings. \textbf{e} Violin plots of the average contextual cross-entropy over $100$ independent runs with different random seeds employing axial strain measurements instead of displacements, showing that accurate contextual inference is maintained across sensing modalities. \textbf{f} Violin plots of the average mechanical performance for the same $100$ strain-measurement runs.}
\label{fig:beam_2}
\end{figure}

The previous results reflect a setting in which the number of available latent context slots coincides with the number of operating regimes, primarily to simplify the presentation of the perception--action dynamics. However, the proposed framework naturally generalizes to settings in which the actual number of operating contexts is a priori unknown. Following recent formulations of structure learning in AIF, the generative model can be initialized with a set of uncommitted ``reserve'' context slots, defining an upper bound on its representational complexity~\cite{smith2020active}. These reserve states become progressively specialized only when required to explain previously unseen patterns of observations. Once a context slot has specialized to represent a recurring operating regime, it continues to encode that regime, while the remaining reserve states remain available to capture novel patterns should they emerge. As a result, previously acquired knowledge is not overwritten, mitigating catastrophic forgetting during adaptation. The idea is conceptually related to Bayesian non-parametric approaches, such as the so-called ``Chinese Restaurant'' process, where model complexity is inferred from data under suitable structural priors~\cite{gershman2012tutorial,stoianov2022hippocampal}. 

To illustrate this capability, the experiment in \fig\ref{fig:beam_2}a overparametrizes the generative model with five latent context slots, despite the presence of only three operating regimes. Initially, these reserve states compete to explain the incoming sensorimotor evidence. As learning proceeds, only the reserve states that consistently explain recurring patterns acquire specialized observation models, whereas the other remain close to their uninformative priors. The Gaussian perturbation added to the initial observation models facilitates this differentiation by breaking the permutation symmetry of the initial reserve states, which would otherwise receive identical Bayesian updates. The overparametrized AIF agent achieves contextual inference accuracy and mechanical performance comparable to those of the correctly specified model (\fig\ref{fig:beam_2}b,~c). Comparable results are also obtained when displacement measurements are replaced by axial strain measurements (\fig\ref{fig:beam_2}e,~f).

After sufficient exploration, BMR is performed at time step 600 (\fig\ref{fig:beam_2}a). This complements online learning and state expansion with an offline model-selection step, in which simpler nested generative models are compared according to their model evidence~\cite{friston2016bayesian,friston2017active}. Candidate reduced models are generated by pruning four levels of the mass-location hidden-state factor. The relative evidence of each candidate model is then evaluated analytically from the posterior Dirichlet concentration parameters accumulated from state occupancies (please see Section~C of the Supplementary Materials for implementation details). The reduced model with the highest evidence is selected to resume the simulation, retaining the learned observation mappings projected onto the simplified state space (\fig\ref{fig:beam_2}d). Because the discarded mass locations correspond to hypotheses that contribute only marginally to explaining the accumulated evidence, the compressed agent exhibits almost identical behavior despite its reduced complexity.

\subsection{Adaptive frame under dynamic excitation}
The second case study leverages the proposed framework to autonomously adapt the dynamic behavior of a structure subjected to changing excitation conditions. Specifically, the agent must tune the stiffness of a frame structure under dynamic loadings to suppress vibrations while using the minimum amount of stiffness resources.

\begin{figure}[t!]
\centering\includegraphics[width=1\linewidth]{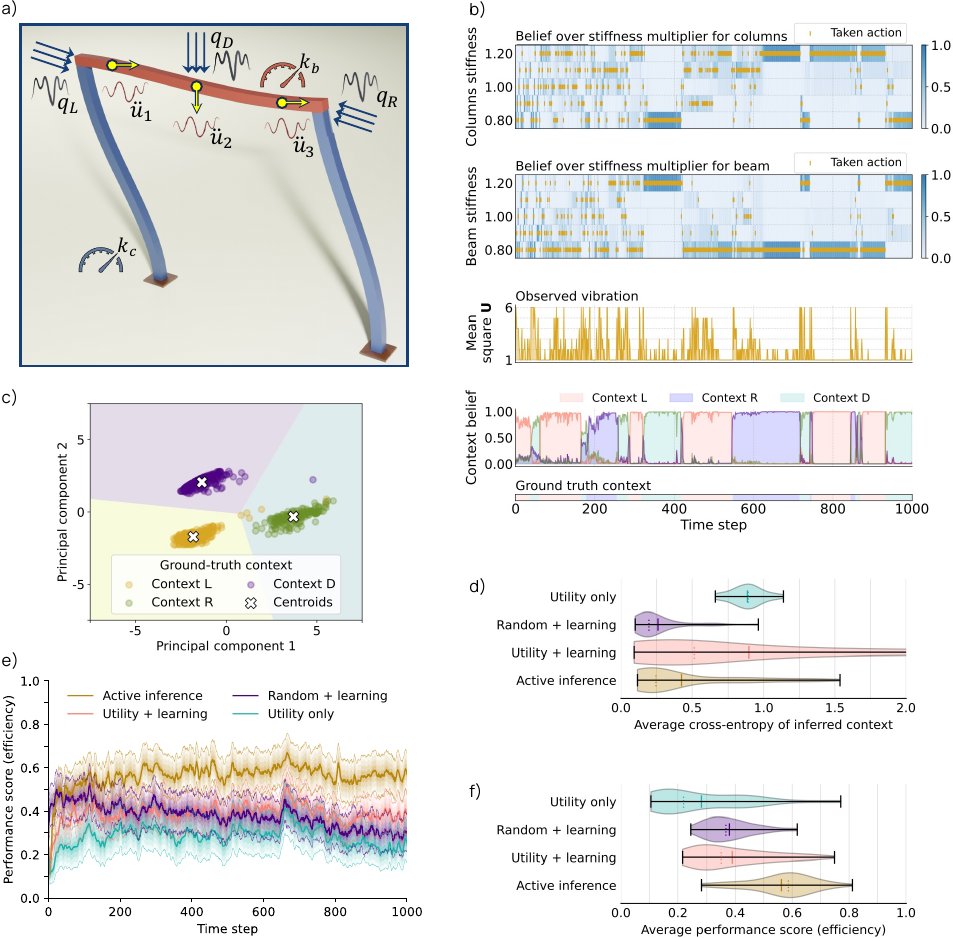}
\caption{\textbf{Case study 2: Adaptive frame for resource-constrained vibration mitigation.}\linebreak\textbf{a} Schematic of the portal frame, showing the three harmonic distributed loads ($q_L,q_D,q_R$), the three accelerometers providing sparse measurements ($\ddot{u}_1,\ddot{u}_2,\ddot{u}_3$) of the structural response, and the tunable stiffness multipliers $k_c$ and $k_b$ associated with the columns and the beam. The operating regime switches stochastically among the three loading contexts (L), (D), and (R). \textbf{b} Representative evolution of the perception--action loop. From bottom to top: ground-truth loading regime; inferred beliefs over the latent loading context; observed vibration level to be minimized; posterior beliefs over the beam and column stiffness multipliers $k_c$ and $k_b$, where color intensity denotes the probability assigned to each stiffness level and the gold marker indicates the selected action. \textbf{c}. Two-dimensional visualization of the neural-reservoir state space obtained by principal component analysis. Each point represents the reservoir embedding of a one-second sensor recording and is colored according to the ground-truth loading context. Background shading denotes the partition induced by k-means clustering, while crosses indicate cluster centroids. \textbf{d} Violin plots of the average cross-entropy between the inferred beliefs and the ground-truth loading regime, quantifying the progressive emergence of contextual understanding for the proposed active inference framework and three reference strategies (utility + learning, random + learning, and utility only) over $50$ independent runs with different random seeds.}
\label{fig:frame_1}
\end{figure}

The system consists of the portal frame shown in \fig\ref{fig:frame_1}a, with columns featuring a $(0.04\times0.04)~\mathrm{m}^2$ cross-section and a height of $1~\mathrm{m}$, and a beam with the same cross-section and a span of $1~\mathrm{m}$. The mechanical properties are those of aluminum, with Young's modulus $70~\mathrm{GPa}$, Poisson's ratio $0.33$, and density $2700~\mathrm{kg/m^3}$. Three operating regimes are generated by harmonic distributed loads applied at the top of the left column $q_L$ (context L), at the midspan of the upper beam surface $q_D$ (context D), and at the top of the right column $q_R$ (context R), respectively. The loads are distributed over an area of $(0.25\times0.04)~\mathrm{m}^2$, oscillate at $10~\mathrm{Hz}$, and have amplitudes corresponding to masses of $100~\mathrm{kg}$ for contexts L and D, and $300~\mathrm{kg}$ for context R. The active loading regime switches stochastically with probability $1/50$ at each physical--digital interaction, producing vibration patterns that are observable only through sparse sensor measurements. 

\setcounter{figure}{4}    
\begin{figure}[t!]
\caption{\textbf{Caption continued.}  \textbf{e} Evolution of the mechanical objective, expressed as the normalized efficiency score $(\mathcal{E}(\mathbf{U})(k_c+k_b))^{-1}$, rewarding vibration suppression and parsimonious use of stiffness, where $\mathcal{E}(\mathbf{U})=\lVert\mathbf{U}\rVert_F^2/(N_lN_s)$ is the mean squared sensor amplitude and $\lVert\,\cdot\, \rVert_F^2$ is the squared Frobenius norm. Solid lines denote averages over the same $50$ runs, and shaded regions indicate $95\%$ confidence intervals. \textbf{f} Violin plots of the average efficiency score.}
\end{figure}

The coupled agent controls two independent stiffness multipliers $k_c$ and $k_b$ associated with the columns and the beam, respectively. Each multiplier is selected from the discrete set $\lbrace0.8,0.9,1.0,1.1,1.2\rbrace$, allowing the agent to continuously adapt the structural dynamics. The control objective requires balancing vibration suppression against the use of stiffness resources in response to the inferred operating conditions.
 
The generative process is obtained by solving the linearized elasto-dynamic problem. The structure is discretized in space using linear tetrahedral finite elements, resulting in $1365$ degrees of freedom. Structural dissipation is modeled through Rayleigh damping, calibrated to provide about $5\%$ damping in the first two vibration modes. The structural response is integrated in time using the implicit Newmark-$\beta$ constant-average-acceleration scheme.

The frame is monitored through $N_s=3$ accelerometers, deployed as shown in \fig\ref{fig:frame_1}a. At each physical--digital interaction, the agent acquires a $1~\mathrm{s}$ recording sampled at $100~\mathrm{Hz}$, yielding multivariate time-series measurements $\mathbf{U}\in\mathbb{R}^{N_l\times N_s}$, with \mbox{$N_l=100$}. After assimilating the measurement window, the agent performs inference, action selection, and learning, before the next interaction.

As in the previous case study, the agent receives two complementary observation modalities. The first consists of clustered reservoir representations obtained by processing the sensor time series through the neural reservoir. These representations encode the spatiotemporal dynamics of the structural response and support contextual inference. The second observation modality is derived from the mean squared amplitude of the measured signals, allowing preferences for vibration suppression. Two additional observation channels provide identity mappings of the selected columns and beam stiffness multipliers, enabling to express preferences for the use of stiffness resources.

The hidden-state factors describe the columns stiffness multiplier, the beam stiffness multiplier, and the latent loading context (see Section~A of the Supplementary Materials for details on the assigned initial prior). As in the previous case study, the observation models are initialized from uninformative Dirichlet priors and progressively refined through Bayesian learning. The transition dynamics is deterministic for the controllable stiffness factors, whereas the latent loading context follows a persistence model that assigns $90\%$ probability to remaining in the current context and distributes the remaining $10\%$ across the alternative contexts.

Preferences over the signal-norm modality encourage vibration suppression, while preferences over the auxiliary stiffness-observation channels promote lower stiffness multipliers. No preferences are assigned to the reservoir-cluster observation modality (please see Section~A of the Supplementary Materials for implementation details and numerical values).

Successful behavior requires the agent to simultaneously infer the active loading regime, discover control policies that suppress structural vibrations while minimizing the amount of stiffness employed, and learn the corresponding observation model. Figure~\ref{fig:frame_1}b demonstrates the emergence of this adaptive behavior. During an initial exploration phase, the agent identifies multiple operating regimes induced by the varying loads and progressively learns how different combinations of beam and column stiffness shape the structural response within each regime. The reservoir encoding transforms continuous sensor measurements into discrete observations that facilitate contextual inference from the underlying spatiotemporal dynamics, despite the continual switching of loading conditions and their coupling with the selected control actions (\fig\ref{fig:frame_1}c). As uncertainty over the latent loading context decreases, the agent transitions from information-seeking to goal-directed behavior. The resulting control strategies are mechanically interpretable: horizontal loading is counteracted by stiffening the columns while maintaining a compliant beam, whereas vertical loading induces the opposite strategy, with a stiffer beam and more compliant columns.

The AIF agent achieves substantially higher contextual inference accuracy than agents lacking epistemic behavior (\fig\ref{fig:frame_1}d). Its cross-entropy between inferred beliefs and the ground-truth loading regime approaches that of the random-exploration agent with online learning, which represents the strongest baseline for contextual representation learning. Mechanical performance is quantified through a normalized efficiency score, inversely proportional to both the observed vibration level and the employed stiffness, as $(\mathcal{E}(\mathbf{U})(k_c+k_b))^{-1}$, where $\mathcal{E}(\mathbf{U})=\lVert\mathbf{U}\rVert_F^2/(N_lN_s)$ is the mean squared sensor amplitude and $\lVert\,\cdot\, \rVert_F^2$ is the squared Frobenius norm. The AIF agent consistently achieves the highest efficiency among all comparison agents (\fig\ref{fig:frame_1}e,~f). In contrast to the previous case study, the increased complexity of the environment prevents the utility-maximization agent with online learning, but without active information seeking, from sporadically converge to a more effective control strategy (please see Section~B of the Supplementary Materials for details on the evaluation metrics computation).

\section{Discussion}\label{sec3}
In this work, we present a computational framework enabling cognitive behavior in mechanical systems by integrating sensing, inference, control, and learning within a closed perception--action loop. The potential of the framework to enable autonomous capabilities in mechanical systems and facilitate the achievement of cognitive devices is demonstrated through numerical simulations. Cognitive mechanical systems equipped with sensing and actuation capabilities are formulated as partially observable dynamical environments~\cite{planning_acting}, coupled with a probabilistic agent that autonomously learns and adapts in pursuit of mechanical objectives through self-directed interaction. Specifically, learning and adaptation emerge from active information gathering that refines the internal representation of the agent. Within this perspective, mechanical systems are conceived not only as assets to be monitored or controlled, but as cognitive systems capable of determining what to sense, what to learn, and what to do. The present work represents an initial step toward such a view of mechanical intelligence, in which structures actively participate in these processes.

The simulation-based case studies used to assess adaptive mechanical behavior involve different sensing modalities and objectives, including static displacement or strain measurements for deformation maximization and dynamic acceleration measurements for resource-constrained vibration mitigation. In all cases, behavior is not driven solely by immediate mechanical utility, as actions are also selected according to their epistemic value in response to uncertainty~\cite{Friston02102015,schwartenbeck2019computational}. In this way, the agent learns online while interacting with the mechanical environment, without separation between data collection, learning, and deployment. Moreover, effective control strategies are obtained with minimal hyperparameter tuning, as all experiments employ the same set of agent and learning parameters. 

A key enabling component is the combination of reservoir computing and AIF, which provide complementary mechanisms for signal processing and probabilistic inference. Reservoir computing encodes sensor measurements while avoiding the expensive training associated with conventional neural networks~\cite{Jaeger,maass2002real,nakajima2021reservoir,yan2024emerging}. This processing is largely agnostic to the physical nature of the sensed quantities. In principle, the same framework could also accommodate acoustic, electrical, thermal, and other sensor modalities. Clustered reservoir representations are used as observations for contextual inference under the AIF generative model~\cite{parr2022active}. These are  further complemented with a signal-norm observation related to the mechanical objective. Because the latter is interpreted conditionally on the inferred context, the same observation space acquires different meanings across operating regimes.

The results also show that only a compact pre-learning phase is required to initialize this process. Rather than relying on large labeled datasets, accumulated experience under arbitrary external inputs and random actions is sufficient to construct the reservoir clusters and estimate the corresponding normalization statistics. This initial phase is deliberately limited, as the agent subsequently acquires knowledge through active interaction with the mechanical environment rather than through extensive passive training on pre-collected data like in contemporary deep learning workflows. Specifically, the agent enters active operation with an uninformative observation model and progressively learns the relationships among observations, actions, and hidden states through Bayesian updates. As uncertainty reduces, behavior gradually transitions from information-seeking to goal-directed, without requiring a mechanistic model of the underlying physical phenomena or external supervision. Beyond enabling autonomous adaptation, the formulation provides an unsupervised approach to structure learning within AIF, allowing the generative model to be constructed and refined directly from raw sensorimotor data. This feature is particularly relevant to engineering applications, where labeled datasets and explicit mappings between latent aspects and observations are often difficult to obtain. Furthermore, inspired by Bayesian non-parametric perspectives on model complexity~\cite{gershman2012tutorial}, adaptation to previously unseen operating regimes is achieved through reserve context states in an overparameterized generative model that can explain newly observed sensorimotor evidence~\cite{smith2020active}. Bayesian model reduction is also integrated to remove redundant hidden-state hypotheses, providing computational savings without substantially affecting control performance. 

Comparisons with alternative agents further highlight the importance of epistemic behavior. Agents deprived of the epistemic component of the expected free energy are less able to reduce uncertainty about the observation model and converge toward less effective control strategies. This suggests that, in uncertain environments, mechanical performance is conditioned not only on learning the consequences of actions, but also on the capability to actively select actions that make the environment more informative~\cite{Friston02102015,schwartenbeck2019computational}. Such a balance between information-seeking and goal-directed behavior is consistent with broader perspectives on embodied intelligence~\cite{pfeifer2007self,liu2025embodied} and emerging digital twins based on autonomous decision-making~\cite{san2026evolution,MT_AIF}.

The proposed framework is therefore distinct from conventional reactive monitoring~\cite{Farrar01,ML_perspective}, where sensing is commonly performed according to predefined acquisition strategies. In our approach, sensing becomes an intrinsic part of an adaptive decision-making process~\cite{friston2010free,buckley2017free}, allowing measurements to be selected based on current uncertainty and their expected contribution to subsequent inference and action. Although active information gathering has been extensively investigated in active perception, robotics, and autonomous navigation~\cite{bajcsy2018revisiting,lanillos2021active,slam}, its integration within mechanical systems remains comparatively unexplored~\cite{andtioris2021102072}. This perspective also highlights a key distinction with respect to reinforcement learning. Indeed, while the considered tasks could, in principle, be addressed via reinforcement learning~\cite{Sutton,kober2013reinforcement,andriotis2021deep}, AIF does not reduce behavior to the maximization of a scalar reward~\cite{friston2017active}. Moreover, reinforcement learning often requires extensive environmental interaction to learn effective policies, which can be impractical for engineering structures. Consequently, state representations, exploration mechanisms, and the treatment of uncertainty differ substantially. Comparing the proposed online setting with alternative paradigms would therefore require reformulating the objectives and learning mechanisms, while introducing an explicit separation between training and deployment. Carrying out such a quantitative comparison remains an interesting direction for future work.

Beyond the numerical demonstrations, the proposed framework suggests unprecedented possibilities toward cognitive materials and structures capable of continual adaptation under evolving conditions, incomplete modeling information, and partial observability. Similar motivations are emerging in collaborative and humanoid robotics, where increasingly sophisticated adaptive behaviors are achieved through substantial computational resources and training~\cite{durr2026outplaying,Lipson_agnostic}. In mechanical systems, however, computational and sensing resources can be considerably more constrained, making the combination of active information gathering, online adaptation, and computationally efficient signal processing particularly relevant. The proposed framework is therefore promising for applications in which mechanistic models are prohibitively difficult to construct, impractical to evaluate, or progressively lose reliability due to long-term degradation, environmental exposure, or sudden operational variations. In structural health monitoring, for instance, civil infrastructures could continuously infer hidden degradation processes from sparse measurements while adapting sensing or response to preserve performance. Similarly, in manufacturing, smart materials and components could autonomously reconfigure their response to maintain precision or suppress undesired vibrations. 

Besides the need for experimental validation of the obtained results, the current study has several limitations that will need to be addressed in future work. First, practical mechanical systems often involve continuous control variables, actuator constraints, and delays, whereas the present study deliberately considers simple discrete action spaces. Similarly, discrete latent states provide an intuitive and computationally tractable representation, but inherently limit the representational capacity of the generative model. Extensions toward continuous AIF formulations~\cite{priorelli2023modeling} will therefore be important for more complex systems. A related challenge concerns the computational scalability of policy inference as the complexity of the generative model increases, which may be addressed by leveraging existing AIF methods based on deep learning and tree search~\cite{fountas2020deep,maisto2021active}. Regarding instead the clustered reservoir representations, clustering and cluster-specific normalization statistics are estimated during pre-learning. This choice provides a simple and efficient interface with the discrete AIF formulation, but limits adaptivity if the operational regimes change substantially. Future implementations could therefore incorporate online adaptation when the existing partition becomes inadequate, for example by triggering re-clustering when free energy exceeds a prescribed threshold. Finally, the context-dependent normalization defining the goal-directed observation introduces a subjective element into the quantification of mechanical performance. Because signal levels are interpreted relative to the inferred dynamical regime, the resulting metric is not an invariant measure. This is not necessarily a limitation, but it highlights the need for other objective performance indicators when comparing with alternative approaches. 

The framework also opens several directions for future methodological developments across both digital and physical domains. One possibility is the integration with adaptive metamaterials~\cite{cummer2016controlling}, whose reconfigurable properties can serve embedded sensing and actuation~\cite{Alu_2025,wu2024wave}. Another is the extension toward distributed systems involving multiple interacting agents~\cite{maisto2023interactive,heins2024collective}, where local perception and action mediated through a shared mechanical substrate could enable collective intelligence without requiring centralized control. A further direction is physical reservoir computing~\cite{nakajima2021reservoir,yan2024emerging,appeltant2011information}, leveraging the inherent nonlinear dynamics of the structure itself as a computational reservoir. Finally, the proposed framework suggests a bidirectional relationship between neuroscience and mechanics. Computational neuroscience provides principles for constructing cognitive digital representations~\cite{Clark2016}, while cognitive mechanical systems could, in turn, serve as simplified platforms where sensing and action can be manipulated to investigate hypotheses about cognition.

\section{Methods}\label{sec4}
In this paper, we develop a framework for cognitive structures that enables mechanical systems to autonomously learn, adapt, and make decisions under partial observability. The following sections describe the key enabling components of the approach. Section ``Neural reservoir'' introduces the reservoir computing architecture for assimilating structural signals and extracting contextual representations. Section ``Active inference'' describes the generative model and the variational free-energy minimization principles underlying perception, action, and learning. Finally, Section ``Integration of the cognitive framework'' describes how these components are coupled to enable adaptive decision-making in partially observable mechanical systems.

\subsection{Neural reservoir}
The neural reservoir is implemented following the echo state network (ESN) formulation~\cite{Jaeger}. It consists of a recurrent network of $N_n=400$ sparsely interconnected artificial neurons. At each step $l$ within a physical--digital interaction, for $l=1,\ldots,N_l$, the reservoir receives a sensor measurements vector $\mathbf{u}_l \in \mathbb{R}^{N_s}$, through an input weight matrix $\mathbf{W}_i \in \mathbb{R}^{N_n \times N_s}$, while internal recurrent interactions are governed by the reservoir matrix $\mathbf{W}_r \in \mathbb{R}^{N_n \times N_n}$. The reservoir state $\mathbf{h}_l\in \mathbb{R}^{N_n}$ evolves as follows:
\begin{equation}
\mathbf{h}_l= f (\mathbf{W}_i\mathbf{u}_l + \mathbf{W}_r\mathbf{h}_{l-1}),
\label{eq:reservoir}
\end{equation}
where $f(\,\cdot\,)$ denotes a nonlinear activation function, here chosen as the hyperbolic tangent. Unlike conventional recurrent neural networks, the reservoir parameters remain fixed and are not optimized during learning. Consequently, the dynamical properties of the reservoir are determined entirely by its initialization.

Following the standard ESN formulation, the entries of both $\mathbf{W}_i$ and $\mathbf{W}_r$ are sampled from a uniform distribution over $[-1,1]$, while sparse connectivity is imposed on $\mathbf{W}_r$ with a connection density of $0.1$. The recurrent matrix $\mathbf{W}_r$ is subsequently rescaled according to its spectral radius $\lambda_\text{max}$, defined as the magnitude of its dominant eigenvalue. In the present work, $\mathbf{W}_r$ is normalized to obtain $\lambda_{\max}=0.9$, as \mbox{$\mathbf{W}r \leftarrow
\frac{0.9}{\lambda{\max}}\mathbf{W}_r$}. This scaling promotes the echo-state property, stabilizing a fading-memory regime in which the influence of initial conditions progressively diminishes.

The reservoir transformation provides a nonlinear representation of spatiotemporal sensor measurements without requiring task-specific training. In contrast to conventional ESN implementations, which typically employ a trained linear readout for downstream supervised tasks, we use the reservoir as a feature map. Specifically, for each acquisition window, we compute a window-level representation by taking the temporal mean of the squared reservoir activations $\mathbf{s}=\frac{1}{N_l}\sum_{l=1}^{N_l}\mathbf{h}_l^{2}$. This representation is subsequently converted into a discrete observation through unsupervised clustering. We employ the mini-batch $k$-means algorithm, with the number of clusters determined by the contextual observation space of the AIF generative model. Each cluster therefore defines a label-free mapping from the continuous representation $\mathbf{s}\in\mathbb{R}^{N_n}$ to a contextual observation $O^{\mathrm{ctx}}\sim p(o^{\mathrm{ctx}})$.

\subsection{Active inference}
Active inference is a computational framework for modeling cognitive processing underlying perception, action, and learning. This triad of fundamental functions is formulated through the minimization of the variational free energy (VFE) functional. To this end, AIF agents are equipped with a generative model that captures the conditional dependencies governing how hidden states generate observations and evolve under the influence of actions. Through this probabilistic representation, the agent interprets observations and evaluates the consequences of actions to perform approximate Bayesian inference of cognitive behavior. The external physical system is instead referred to as the generative process, i.e., the physical process that generates sensor measurements through which the system becomes partially observable to the agent.

\subsubsection{Generative model}
\begin{figure}[!t]
\centering\includegraphics[width=0.7\linewidth]{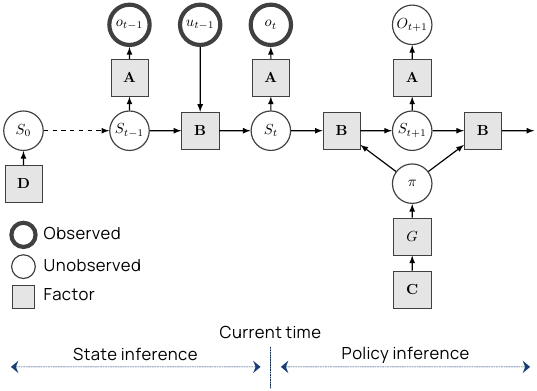}
\caption{\textbf{Generative model of a partially observable Markov decision process.} Circular nodes represent random variables. Nodes with bold outlines indicate observed quantities, while those with thin outlines represent latent variables to be inferred. Gray square nodes represent parametrized components of the generative model. Directed edges indicate conditional dependencies parametrized by these factors. The $\mathbf{A}$ factor parametrizes the observation model, i.e., the likelihood describing how hidden states $S_t$ generate observations $O_t$. The $\mathbf{B}$ factor parametrizes the transition model, i.e., the likelihood describing how hidden states $S_{t-1}$ evolves into $S_{t}$ conditioned on action $U_{t-1}$ under policy $\pi$. The $\mathbf{C}$ factor encodes prior preferences over observations and directly influences the $G$ factor associated with the expected free energy objective that governs policy inference. The $\mathbf{D}$ factor encodes the initial prior over hidden states, representing the agent's belief before any observations are incorporated. Hidden-state inference assimilates observations collected up to the current time. Policy inference evaluates future actions, propagating the updated beliefs forward in time.}
\label{fig:graph}
\end{figure}

Figure~\ref{fig:graph} illustrates the factor graph representing the generative model, formulated as a dynamic Bayesian network~\cite{koller2009probabilistic}. The abstraction is inspired by classical POMDP formulations~\cite{2010artificial}. Circular nodes represent random variables at discrete time steps. Nodes with bold outlines indicate observed quantities, whereas those with thin outlines represent latent variables to be inferred. Directed edges indicate conditional dependencies parametrized by the components of the generative model, represented by gray square nodes.

Capital letters denote random variables associated with the quantities in our abstraction and the corresponding lowercase letters refer to their realizations. Subscripts indicate time indices, where $t=0$ marks the beginning of active operation. Calligraphic letters denote the sets of possible values that each quantity can assume. For instance, hidden states are denoted as $S_t\sim p(s_t)$, where $s_t$ represents a particular realization at time $t$, and $p(s_t)$ defines the probability that $S_t=s_t$ for any possible state $s_t\in \mathcal{S}$. The state space $\mathcal{S}$ can represent initial and/or boundary conditions, material properties, or any other physical feature relevant for prediction and decision-making~\cite{Ferrari2024}.

Probabilistic inference of hidden states, i.e., perception, is mediated by the assimilation of observations $O_t \sim p(o_t)$. The observation space $\mathcal{O}$ may include sensor measurements, inspection results, or pre-processed information such as the context-related observation from reservoir encoding. This perceptual process is realized through observation models -- one for each observation modality -- which model observations as being stochastically generated from hidden states.

The posterior belief over hidden states guides the selection of control actions to influence future physical states. In \fig\ref{fig:graph}, $U_t\sim p(u_t)$ denotes a decision variable from the action space $\mathcal{U}$. Actions may affect the mechanical properties of the system, adjust the operational conditions, or modify the sensing process. Each action is associated with a transition model -- one for each hidden-state factor -- defining a control-dependent predictor that propagates beliefs forward in time. The control variable is selected according to a policy $\pi$, which defines a sequence of control actions. Policies are treated as latent variables to be inferred, whose posterior encodes the agent's beliefs about the actions to take.

The parametrized factors defining the generative model include the observation model $\mathbf{A}:\mathcal{O}\times\mathcal{S}\mapsto[0,1]$, which encodes the sensory likelihood $p(O_t\mid S_t;\boldsymbol{\phi})$; the transition model $\mathbf{B}:\mathcal{S}\times\mathcal{S}\times\mathcal{U}\mapsto[0,1]$, which encodes the Markovian evolution of hidden states conditioned on control actions $p(S_t\mid S_{t-1},U_{t-1};\boldsymbol{\phi})$; and the initial prior $\mathbf{D}:\mathcal{S}\mapsto[0,1]$, which encodes the initial belief $p(S_0;\boldsymbol{\phi})$ before any observations are incorporated. The preferences $\mathbf{C}:\mathcal{O}\mapsto[0,1]$ encode utility-maximization goals as a prior distribution over future observations $\widetilde{p}(O_{t_c:t_p})$ from the current time $t_c$ to a prediction horizon $t_p > t_c$. This prior may represent real costs associated with states and actions, or abstract metrics tuned to steer the system toward the desired behavior. It directly influences the expected free energy (EFE) factor $G$, which quantifies the desirability of each policy.

Relying on a discrete POMDP formulation, hidden states, observations, and actions are defined over discrete and finite spaces. The corresponding conditional probability distributions are represented as categorical distributions, whose parameters are assigned conjugate Dirichlet priors. Learning of the generative model is then formulated through closed-form Bayesian updates of the associated Dirichlet concentration parameters $\boldsymbol{\phi}$. Discrete generative models have been implemented using \texttt{pymdp}~\cite{pymdp}.

\subsubsection{Perception}
At the current time step $t_c$, given an observation $O_{t_c}=o_{t_c}$, an hidden state belief $S_{t_c-1}$ at the previous time step, and an action $U_{t_c-1}=u_{t_c-1}$ selected according to policy $\pi$, the joint belief state from $t=0$ to $t=t_c$, with $t= 0,\ldots,t_c$, can be factorized by leveraging the conditional dependencies implied by the graph structure, as follows:
\begin{equation}
\begin{split}
&p(S_{0:t_c}, U_{0:{t_c-1}}, O_{1:t_c}, \boldsymbol{\phi}) =\\&\hspace{50pt} p(\boldsymbol{\phi})p(S_0;\boldsymbol{\phi})\prod_{t=1}^{t_c}p(O_t\mid S_t;\boldsymbol{\phi})p(S_t\mid S_{t-1},U_{t-1};\boldsymbol{\phi})p(U_{t-1}),
\end{split}
\label{eq:joint_factor}
\end{equation}
where $p(\boldsymbol{\phi})$ represents the prior distribution over the model parameters $\boldsymbol{\phi}$.

Since exact Bayesian inference over dynamically evolving hidden states is generally intractable, AIF resorts to variational inference~\cite{murphy2023probabilistic}. Given a tractable variational distribution $Q(S_{t_c};\boldsymbol{\theta}):\mathcal{S}\mapsto[0,1]$, parametrized by variational parameters $\boldsymbol{\theta}$, the posterior belief is approximated by minimizing the VFE, as follows:
\begin{equation}
\begin{split}
\boldsymbol{\theta}^* &= \argmin_{\boldsymbol{\theta}}\ F_{t_c}(\boldsymbol{\theta})\\
&=\argmin_{\boldsymbol{\theta}}\ \mathbb{E}_{Q(S_{t_c})}\left[\ln{Q(S_{t_c}; \boldsymbol{\theta})}-\ln{p(o_{t_c}, S_{t_c}\mid S_{t_c-1},u_{t_c-1};\boldsymbol{\phi})}\right]\\
&=\argmin_{\boldsymbol{\theta}}\ \mathbb{E}_{Q(S_{t_c})}\left[\ln{Q(S_{t_c}; \boldsymbol{\theta})}-\ln{\left(p(o_{t_c}\mid S_{t_c};\boldsymbol{\phi})p(S_{t_c}\mid S_{t_c-1},u_{t_c-1};\boldsymbol{\phi})\right)}\right],
\end{split}
\label{eq:inference}
\end{equation}
where $\mathbb{E}_{Q(S_{t_c})}$ denotes the expectation with respect to the variational posterior. The variational parameters $\boldsymbol{\theta}$ define the categorical probabilities associated with the possible states in $\mathcal{S}$, and \eq\eqref{eq:inference} yields the hidden-state posterior that best matches the incoming observations with the learned generative model.

The posterior $Q(S_{t}; \boldsymbol{\theta})$ is further factorized across $F$ independent hidden-state factors $S=\lbrace S^1,\ldots,S^F\rbrace$, representing distinct aspects of the generative process, as $Q(S_{t}; \boldsymbol{\theta})=\prod_{f=1}^FQ(S^f_{t}; \boldsymbol{\theta})$. Similarly, observations are structured into $M$ modalities \mbox{$O=\lbrace O^1,\ldots,O^M\rbrace$}, corresponding to the sensing modalities available to the agent.

In this multi-modal, multi-factor formulation, the observation likelihood is represented by a collection of $M$ arrays $\mathbf{A}=\lbrace \mathbf{A}^1,\ldots,\mathbf{A}^M\rbrace$, defining the observation model for each modality. Similarly, the transition model is represented by a collection of $F$ arrays $\mathbf{B}=\lbrace \mathbf{B}^1,\ldots,\mathbf{B}^F\rbrace$. Control variables are factorized analogously to hidden states, such that $U=\lbrace U^1,\ldots,U^F\rbrace$. Each control factor governs the transitions of the corresponding hidden-state factor, under the assumption that these evolve independently.

The marginal variational posteriors for each hidden state factor are computed via mean-field fixed-point iteration~\cite{wainwright2008graphical}. The algorithm proceeds by setting the gradient of the VFE with respect to the variational parameters to zero, and iteratively solving for each factorized component $Q(S^f_{t_c}; \boldsymbol{\theta})$, with $f=1,\ldots,F$. A detailed derivation of this procedure is provided in~\cite{pymdp}.

\subsubsection{Action}
Given the updated hidden-state posterior, the desirability of each admissible policy is measured through the EFE, a quantity that balances three complementary aspects: \textit{curiosity}, \textit{saliency}, and \textit{utility}. Curiosity favors actions that steer toward underexplored regimes, allowing the generative model to progressively improve by learning the statistical regularities characterizing the environment. Saliency favors actions that perturb the environment to acquire observations that reduce uncertainty about hidden states. Utility favors actions that produce observations consistent with preferred outcomes (associated with the mechanical objectives). Curiosity and saliency therefore promote epistemic behavior through uncertainty resolution, whereas utility drives goal-directed behavior toward prior preferences.

Like the VFE, the EFE depends on observations and hidden states. However, while the VFE quantifies the discrepancy between the agent's beliefs and the observations currently available, the EFE requires computing expectations over future hidden states and observations predicted by the generative model under each candidate policy.

Selecting the policy with the lowest EFE enables adaptive decision-making without an explicit separation between training and deployment. Specifically, policies are sampled according to the following posterior distribution:
\begin{equation}
p(\pi) = \sigma(-\gamma\mathbf{G}),
\label{eq:EFE_prior}
\end{equation}
where $\mathbf{G}=(G^{\pi_1},\ldots,G^{\pi_P})^\top\in\mathbb{R}^P$ assigns an EFE to each of the $P$ feasible policies $\Pi=\lbrace\pi_1,\ldots,\pi_P\rbrace$, $\sigma(\,\cdot\,)$ denotes the Softmax function, and $\gamma\in\mathbb{R}^+$ is an inverse-temperature parameter. Higher $\gamma$ values yield increasingly deterministic policy selection.

As adapted from~\cite{pymdp}, the EFE at a generic time step $t=t_c,\ldots,t_p$ for policy $\pi$ is given by:
\begin{equation}
\begin{split}
G_t^\pi&=-\underbrace{\mathbb{E}_{Q(O_t\mid\pi)}[\text{D}_\text{KL}[Q(\boldsymbol{\phi}\mid O_{t},\pi)\mid\mid Q(\boldsymbol{\phi}\mid\pi)]]}_{\text{Epistemic value (curiosity)}}\\&\hspace{10pt}-\underbrace{\mathbb{E}_{Q(O_t\mid\pi)}[\text{D}_\text{KL}[Q(S_{t}\mid O_{t},\pi)\mid\mid Q(S_{t}\mid\pi)]]}_{\text{Epistemic value (saliency)}}-\underbrace{\mathbb{E}_{Q(O_t\mid\pi)}[\ln{\widetilde{p}(O_t)}]}_{\text{Utility}}\\&\hspace{10pt}+\underbrace{\mathbb{E}_{Q(O_t\mid\pi)}[\text{D}_\text{KL}[Q(S_{t},\boldsymbol{\phi}\mid O_{t},\pi)\mid\mid p(S_t,\boldsymbol{\phi}\mid O_t,\pi)]]}_{\text{Expected approximation error ($\geq0$)}},
\end{split}
\label{eq:EFE}
\end{equation}
where $\text{D}_\text{KL}\left[\,\cdot\,\right]$ denotes the Kullback-Leibler divergence and, for simplicity, we omit the dependence of the variational posterior on $\boldsymbol{\theta}$. The first term captures the epistemic value associated with the expected information gain about the parameters governing the generative model. It favors policies that are expected to produce informative observations for refining the agent's internal model. The second term captures the epistemic value associated with the expected information gain about hidden states. These two terms quantify the expected change in beliefs induced by future observations, allowing the agent to anticipate how informative each candidate policy is expected to be for future performance. The third term instead corresponds to goal-directed utility, as measured by the expected log probability of the preferred observations. It therefore favors policies expected to generate outcomes aligned with the agent's prior preferences. The final term represents the expected approximation error associated with the variational approximation; following~\cite{pymdp}, it is assumed to be negligible and omitted in the implementation. For policies extending over multiple time steps, the EFE is obtained by summing the time-specific contributions $G^\pi=\sum_{t=t_c}^{t_p}G_t^\pi$.

\subsubsection{Learning}
The stochastic parametrization underlying each categorical distribution is obtained by decomposing $\boldsymbol{\phi}$ into subsets corresponding to the Dirichlet parameters associated with the arrays $\mathbf{A}$, $\mathbf{B}$, and $\mathbf{D}$. In what follows, we focus on learning the parameters of $\mathbf{A}$, although the same reasoning can be applied to $\mathbf{B}$ and $\mathbf{D}$. Online learning of these parameters is a slower-timescale updating process compared with the faster inference for hidden states and policies. This is enabled by the conjugate-prior formulation, which allows closed-form Bayesian updates.

The parametrization of the observation model $\mathbf{A}\in\mathbb{R}^{\mid\mathcal{O}\mid\times\mid\mathcal{S}\mid}$, encoding the sensory likelihood $p(O_t\mid S_t;\mathfrak{A})$, is given by the tensor of categorical parameters \mbox{$\mathfrak{A}\in\mathbb{R}^{\mid\mathcal{O}\mid\times\mid\mathcal{S}\mid}$}, such that
\begin{align}
O_t\mid S_{t};\mathfrak{A}&\sim\text{Cat}(\mathfrak{A}),\\
p(\mathfrak{A})&=\prod_{s\in\mathcal{S}} p(\mathfrak{A}_{\bullet,s}),\quad \mathfrak{A}_{\bullet,s}\sim\text{Dir}(\mathfrak{a}_{\bullet,s}),
\end{align}
where $\mid\cdot\mid$ denotes set cardinality; $\text{Cat}(\,\cdot\,)$ and $\text{Dir}(\,\cdot\,)$ denote categorical and Dirichlet distributions, respectively; the notation $\mathbf{X}_{\bullet,j}$ refers to the $j$th column of a generic matrix $\mathbf{X}$; and $\mathfrak{a}  \in\mathbb{R}^{\mid\mathcal{O}\mid\times\mid\mathcal{S}\mid}$ collects the positive concentration parameters defining the Dirichlet prior over $\mathfrak{A}$. For notational simplicity, we assume the generative model is not factorized into multiple hidden-state factors or observation modalities.

Learning is formulated as approximate inference over $\mathfrak{A}$ by minimizing the VFE with respect to its variational posterior $Q(\mathfrak{A})$, which is modeled as follows:
\begin{align}
Q(\mathfrak{A})&=\prod_{s\in\mathcal{S}} Q(\mathfrak{A}_{\bullet,s}), &\mathfrak{A}_{\bullet,s}&\sim\text{Dir}(\widehat{\mathfrak{a}}_{\bullet,s}),
\end{align}
where $\widehat{\mathfrak{a}}\in\mathbb{R}^{\mid\mathcal{O}\mid\times\mid\mathcal{S}\mid}$ plays the same role as $\mathfrak{a}$ in parametrizing the Dirichlet distribution, while being treated as the variational parameters of the posterior.
  
Following~\cite{pymdp}, given the Dirichlet prior parameters $\mathfrak{a}$, the observation $O_{t_c}=o_{t_c}$, and the hidden state posterior $Q^*(S_{t_c})$, the fixed-point update rule for the variational posterior Dirichlet parameters $\widehat{\mathfrak{a}}$ is:
\begin{equation}
\widehat{\mathfrak{a}}^*=\mathfrak{a}+\eta(o_{t_c}\otimes Q^*(S_{t_c})),
\label{eq:Dirichlet}
\end{equation}
where $\otimes$ denotes the outer product and $\eta\in\mathbb{R}$, with $0\leq\eta\leq1$, is a learning rate parameter that scales the update step.

\subsection{Integration of the cognitive framework}
The ESN and AIF components are integrated within a closed perception--action loop that couples a cognitive mechanical system with its digital representation. The physical system is represented by a parametrized finite element model of an elasto-dynamic problem, solved using \texttt{DOLFINx}~\cite{dolfin}. This virtual environment provides a numerical proxy for the generative process, which evolves under external inputs and selected actions while generating the sensor measurements through which the mechanical system is partially observable.

At each interaction, the generative process is advanced over a finite acquisition window under the current environment conditions and control actions. The resulting sensor measurements are collected as the multivariate time series $\mathbf{U}$. The dynamics is propagated continuously across successive interactions, such that each acquisition window inherits the mechanical state resulting from the preceding interactions.

Before active operation, the mechanical system undergoes a pre-learning phase of $N_i=1000$ interactions under arbitrary external inputs and randomly selected actions. The resulting sensor recordings are collected as  $\lbrace\mathbf{U}_i\rbrace_{i=1}^{N_i}$ and standardized to display zero mean and unit variance along each channel. Each acquisition window is then processed through a fixed ESN, using \eq\eqref{eq:reservoir} with the reservoir state reset to zero at the beginning of each interaction. For the $i$th recording, $i=1,\ldots,N_i$, the resulting sequence of reservoir activations is aggregated into the window-level representation $\mathbf{s}_i$, as described in Section ``Neural reservoir''. The collection of these representations $\lbrace\mathbf{s}_i\rbrace_{i=1}^{N_i}$ is used to fit a mini-batch $k$-means algorithm. The number of clusters $N_c$ is selected according to the cardinality of the contextual observation space of the generative model, which coincides with the number of context slots available to the agent. The resulting cluster centroids serve to partition the dynamical regimes during active operation. In addition, the pre-learning recordings, assigned to their respective clusters, serve to estimate cluster-specific normalization statistics for the signal-norm observation modality $O^\mathcal{E}\sim p(o^\mathcal{E})$. Specifically, we compute the $10$th and $90$th percentiles from the mean squared sensor amplitudes $\mathcal{E}(\mathbf{U}_i)$ of the recordings assigned to cluster $c$, respectively denoted by $q_{10}^c$ and $q_{90}^c$, for $c=1,\ldots,N_c$. These contextual normalization statistics allow the same discrete observation space to represent signal magnitudes (and the associated preferences) across different operating regimes.

The agent is initialized with a flat observation model, perturbed by zero-mean Gaussian noise with variance $5\cdot10^{-4}$. The number of interactions during active operation is selected case-by-case to display convergence of the learned observation model and achieve qualitatively stable behavior. Each interaction begins with the acquisition of sensor measurements from the generative process. The measurements are standardized and subsequently processed by the fixed ESN. The resulting reservoir trajectory is aggregated into the window-level representation and clustered, providing the contextual observation $o^{\mathrm{ctx}}=\argmax_c\ p(c\mid\mathbf{s})$, $c=1,\ldots,N_c$, with clusters probabilities obtained from the normalized inverse distances between $\mathbf{s}$ and the cluster centroids. The signal-norm observation is instead computed by normalizing the mean squared sensor amplitude using the statistics associated with the inferred context and then discretizing into $N_g$ ordered levels, as $o^\mathcal{E} = \left\lceil  N_g \ \text{clip}(\frac{\mathcal{E}(\mathbf{U})-q_{10}^c}{q_{90}^c-q_{10}^c},0,1)\right\rceil$, where $\lceil\,\cdot\,\rceil$ denotes the ceiling function. All the experiments presented in Section ``Results'' rely on $N_g=6$ discretization levels. 

The agent assimilates the resulting discrete observations, together with the actions from the preceding interaction, to update its posterior beliefs over hidden states through the variational inference \eq\eqref{eq:inference}. The updated beliefs are then propagated through the generative model to evaluate the EFE of ``what-if'' trajectories associated with the admissible policies. In the present implementation, policies have a temporal depth of one time step and are evaluated by computing the epistemic and utility contributions in \eq\eqref{eq:EFE}. The inverse-temperature parameter scaling the precision of the posterior over policies in \eq\eqref{eq:EFE_prior} is set to $\gamma=16$. The sampled action is subsequently applied to the mechanical system, which is advanced through the next acquisition window closing the perception--action loop. As interactions accumulate, the agent progressively adapts the Dirichlet parameters of the observation model through online Bayesian learning, using \eq\eqref{eq:Dirichlet} with a learning rate of $\eta=0.6$.

\backmatter

\subsection*{Data availability}
No additional data, apart from the provided implementation code, are necessary to reproduce the simulations detailed in Section ``Results''.

\subsection*{Code availability}
The implementation code of the proposed methodology is publicly available in the \texttt{AIFmech} repository~\cite{AIFmech} at \url{https://github.com/MatteoTorzy/AIFmech}. The code can be used to reproduce the simulations and the plots presented in Section ``Results''.


\subsection*{Acknowledgements}
This work is supported by the ERC advanced grant IMMENSE (Grant Agreement 101140720), funded by the European Union. Views and opinions expressed are however those of the authors only and do not necessarily reflect those of the European Union or the European Research Council Executive Agency. Neither the European Union nor the granting authority can be held responsible for them. Author AM also acknowledges the financial support from the FIS starting grant DREAM (Grant Agreement FIS00003154), funded by the Italian Science Fund (FIS) - Ministero dell’Università e della Ricerca.  
Authors DM, FD, and GP acknowledge financial support from the ERC consolidator grant ThinkAhead (Grant Agreement 820213), funded by the European Union.

\subsection*{Author contribution}
MT, GP, and AC conceptualized the study, with contributions from DM, AM, and FD. MT conceived the methodology and developed the software implementation. MT implemented the test cases and performed the analysis. MT wrote the manuscript. All authors contributed to reviewing and editing the manuscript. AC supervised the work.

\subsection*{Competing interests} 
The authors declare no competing interests.

\subsection*{Additional information} 
\begin{itemize}
\item \textbf{Correspondence.} Correspondence and requests for materials should be addressed to Matteo Torzoni.
\end{itemize}

\newpage
\renewcommand{\thesection}{\Alph{section}.}
\renewcommand{\theequation}{S.\arabic{equation}}
\renewcommand{\thepage}{S~--~\arabic{page}}
\setcounter{page}{1}
\setcounter{equation}{0}

\begin{center}
\Titlefont{Supplementary materials}
\end{center}
\bigskip

\numbered
\section{Assigned initial beliefs and prior preferences}

\paragraph{Initial prior over hidden states}

The initial prior $\mathbf{D}$ over hidden states is specified through strongly peaked distributions for each hidden-state factor. The corresponding initial belief vectors are obtained by applying a Softmax transformation to one-hot vectors with amplitude $100$, resulting in an almost deterministic initial belief over the selected states. For hidden-state factors with an explicit semantic interpretation, the selected state corresponds to the known initial configuration of observable aspects of the mechanical system, reflecting the assumption that the agent and the actuation mechanism are coordinated. Specifically, in the first case study, the initial belief over the mass position is concentrated on the leftmost position $0.05~\mathrm{m}$. In the second case study, the initial beliefs over the two controllable stiffness multipliers are both concentrated on the central value $1.0$.

A different interpretation applies to the latent context factors, whose semantic meaning is not specified a priori but emerges through interaction and learning. For these factors, the same strongly concentrated initialization is used consistently across simulations, but the selected state does not correspond to any particular physical regime. While the agent is initialized as being in a definite latent context, it has no prior knowledge of the relationship between this context and the operational conditions that generate the observations. The arbitrary choice of the initially active context state therefore has neither predefined semantic meaning nor a prescribed effect on the simulation results.

\paragraph{Prior preferences over observations}
 
The prior $\mathbf{C}$ over observations, or prior preferences, represents abstract metrics designed to steer behavior toward the mechanical objectives of the two case studies. These preferences are factorized across observation modalities, with the preference associated with each outcome expressed as a relative log-probability. The resulting log-probability vectors are passed through a Softmax function to produce valid probability distributions, which are then used to compute the expected utility term of the expected free energy for policy inference. The relative magnitudes of the preference terms determine their relative contribution to action selection. No additional weighting or normalization is introduced between heterogeneous quantities in the current implementation, such that all modalities are assigned an implicit unit weight.

In both case studies, no preferences are assigned to the reservoir-cluster observation modality $O_t^{\mathrm{ctx}}\sim p(o_t^{\mathrm{ctx}})$, as encoded using uniform log-probabilities. This modality therefore serves exclusively to provide information for contextual inference and remains preference-free. 

For the first case study, preferences are specified over the signal-norm modality $O_t^\mathcal{E}\sim p(o_t^\mathcal{E})$ to promote beam deformation, as follows:
\begin{equation}
\mathbf{C}^\mathcal{E} \leftarrow [\ln{\widetilde{p}(O^\mathcal{E}_{t})}]_i = i-1,
\end{equation}
where $i=1,\ldots,N_g$, with $N_g=6$, denotes the discretized signal-norm observation level, and the modality ordering is defined such that increasing observation indices correspond to larger deformation levels. According to the deformation-maximization objective, this assigns progressively stronger preferences to higher mean squared sensor amplitudes $\mathcal{E}$.
       
For the second case study, a similar preference structure is used to favor lower vibration levels through the discretized signal-norm observation $O_t^\mathcal{E}$, while additional preferences are assigned to the two auxiliary observation modalities associated with the column and beam stiffness multipliers $O_t^{k_c}\sim p(o_t^{k_c})$ and $O_t^{k_b}\sim p(o_t^{k_b})$, respectively. Collectively, we encode them as follows:
\begin{align}
\mathbf{C}^\mathcal{E} &\leftarrow [\ln{\widetilde{p}(O^\mathcal{E}_{t})}]_i = -(i-1)^{1.3},\\
\mathbf{C}^{k_c} &\leftarrow [\ln{\widetilde{p}(O^{k_c}_{t})}]_j  = -(j-1)^{0.6},\\
\mathbf{C}^{k_b} &\leftarrow [\ln{\widetilde{p}(O^{k_b}_{t})}]_k  = -(k-1)^{0.6},
\end{align}
where $i=1,\ldots,N_g$, with $N_g=6$, denotes the discretized signal-norm observation level and $j,k=1,\ldots,5$ denote the discretized stiffness levels. These preferences favor lower mean squared sensor amplitudes $\mathcal{E}$ and lower stiffness multipliers $k_c,\,k_b$, respectively, thereby reflecting the competing performance and resource objectives through a resource constraint associated with the use of structural stiffness.

\section{Evaluation metrics}
The metrics quantifying the performance of the agents and used to generate the plots in the main text are evaluated independently for each random seed and subsequently aggregated.

\paragraph{Performance score}
For the first case study, the performance score stored at each physical--digital interaction is directly obtained from the signal-norm observation $O_t^\mathcal{E}$, which is computed by normalizing the mean squared sensor amplitude $\mathcal{E}$ using the statistics associated with the corresponding inferred context and then discretizing the normalized value into $N_g$ ordered levels, as \mbox{$o_t^\mathcal{E} = \left\lceil  N_g \ \text{clip}(\frac{\mathcal{E}-q_{10}^c}{q_{90}^c-q_{10}^c},0,1)\right\rceil$}, $c=1,\ldots,N_c$, where $\lceil\,\cdot\,\rceil$ denotes the ceiling function, $\text{clip}(\,\cdot\,,0,1)$ restricts its argument to the interval $[0,1]$, and $N_c$ is the total number of clusters. All the experiments presented in the main text rely on $N_g=6$ discretization levels. 

The time-averaged performance score over a simulation is computed as follows:
\begin{equation}
J_\mathcal{E} = \frac{1}{T}\sum_{t=1}^{T} o_t^\mathcal{E},
\end{equation}
where $T$ denotes the number of physical--digital interactions. The distribution of $J_\mathcal{E}$ across $N^{(1)}_r=100$ runs with different random seeds is used to generate the violin plots reported in Figures~3 and~4 of the main text.

For the temporal performance analysis, the performance score is instead averaged across seeds at each time step, as follows:
\begin{equation}
\overline{o^\mathcal{E}}_t = \frac{1}{N^{(1)}_r}\sum_{n=1}^{N^{(1)}_r} [o_t^\mathcal{E}]_n,
\end{equation}
where $[o_t^\mathcal{E}]_n$ denotes the performance score obtained for seed $n$ at time step $t$. The associated standard error of the mean (SEM) is computed as follows:
\begin{equation}
\mathrm{SEM}(o_t^\mathcal{E}) = \frac{\operatorname{std}_n([o_t^\mathcal{E}]_n)}{\sqrt{N^{(1)}_r}},
\end{equation}
where $\operatorname{std}_n(\,\cdot\,)$ denotes standard deviation across runs. Both the mean trajectory and its SEM are smoothed using a moving average with a window of $6$ time steps. The temporal evolution plot reported in Figure~3 of the main text displays the smoothed mean together with uncertainty bands extending up to $\pm 2\,\mathrm{SEM}$.

\paragraph{Efficiency score}
For the second case study, performance is characterized by an efficiency score that rewards vibration suppression and parsimonious use of structural stiffness. Let $o_t^\mathcal{E}$ denote the vibration-related signal-norm observation and let $o_t^{k}=o_t^{k_c}+o_t^{k_b}$ denote the total actuation effort at time step $t$, i.e., the sum of the actuation measures associated with the columns and beam stiffness. The instantaneous efficiency score is defined as:
\begin{equation}
\eta_t = \frac{(o_t^\mathcal{E}o_t^{k})^{-1} - \eta^\text{min}}{\eta^\text{max} - \eta^\text{min}},
\end{equation}
which corresponds to a min--max scaled score, with extreme values \mbox{$\eta^\text{min} = 0.07$} and $\eta^\text{max} = 0.625$. The corresponding time-averaged efficiency for an individual simulation is $J_{\eta} = \frac{1}{T}\sum_{t=1}^{T}\eta_t$. The violin plot reported in Figure~5 of the main text shows the distribution of $J_{\eta}$ across $N^{(2)}_r=50$ runs with different random seeds.

For the temporal efficiency analysis, the instantaneous value is averaged across seeds, as \mbox{$\overline{\eta}_t =\frac{1}{N^{(2)}_r}\sum_{n=1}^{N^{(2)}_r}[\eta_t]_n$}, with corresponding $\mathrm{SEM}(\eta_t) =\frac{\operatorname{std}_n([\eta_t]_n)}{\sqrt{N^{(2)}_r}}$. As for the first case study, both quantities are smoothed using a moving-average window of $6$ time steps. The temporal evolution plot reported in Figure~5 of the main text displays the smoothed mean together with uncertainty bands extending up to $\pm 2\,\mathrm{SEM}$.

\paragraph{Context inference accuracy}
The inference accuracy for the hidden operating context is quantified through the cross-entropy between the inferred posterior distribution and the ground-truth physical context. Since hidden states do not have a predefined semantic correspondence with the physical contexts, the inferred posterior distribution must first be aligned with the ground-truth context classes. This alignment is performed a posteriori by computing the optimal correspondence between inferred latent states and physical contexts using a maximum-overlap assignment based on the Hungarian algorithm. Specifically, for each simulation span, the  maximum-a-posteriori latent state is first extracted at every time step. A confusion matrix is then constructed between this best-point estimate and the ground-truth context labels to determine the hidden-states permutation that maximizes correspondence with the ground-truth classes. The Hungarian assignment is recomputed separately for each seed to ensure invariance to permutations of the learned latent-state labels.

After alignment, posterior probabilities associated with different hidden states assigned to the same ground-truth class are summed and normalized. The resulting posterior $Q^\text{ctx}_t(\ell)$, $\ell=1,\ldots,N_\text{ctx}$, defines the inferred probability assigned to the physical context $\ell$ at time step $t$, with $N_\text{ctx}$ being the number of ground-truth physical contexts. Therefore, the time-averaged contextual cross-entropy for an individual simulation is defined as follows:
\begin{equation}
J_\mathcal{H}=-\frac{1}{T} \sum_{t=1}^{T} \log Q^\text{ctx}_t(\ell^\text{GT}_t),
\end{equation}
where $\ell^\text{GT}_t$ denotes the ground-truth context at time step $t$. The distribution of $J_\mathcal{H}$ across seeds is used to generate the violin plots reported in Figures~3, 4, and~5 of the main text.

For the temporal evolution analysis, the instantaneous cross-entropy is averaged across seeds, as $\overline{\mathcal{H}}_t =\frac{1}{N_r}\sum_{n=1}^{N_r}[-\log Q^\text{ctx}_t(\ell^\text{GT}_t)]_n$, with corresponding \mbox{$\mathrm{SEM}(\mathcal{H}_t) =\frac{\operatorname{std}_n([\mathcal{H}_t]_n)}{\sqrt{N_r}}$}. These quantities are smoothed using the same six-step moving-average window used for the performance metrics, and used to generate the temporal evolution plot reported in Figure~3 of the main text, displaying the mean together with uncertainty bands extending up to $\pm 2\,\mathrm{SEM}$.

\paragraph{Observation-model entropy}
The uncertainty of the learned observation model is quantified through the normalized Shannon entropy of the likelihood distributions encoded in the observation matrices associated with $M$ observation modalities. For the $m$th observation modality, \mbox{$m=1,\ldots,M$}, the observation model $A^m$ defines the conditional probability distribution $p(O^m_t\mid S_t)$. Given the modality-specific observation model $A^{m}_t(o\mid s)$ at time step $t$, the entropy associated with a hidden-state realization $s$ is defined as follows:
\begin{equation}
\mathcal{L}^{m}_t(s) = -\sum_{o=1}^{N_m} A^{m}_t(o\mid s) \log A^{m}_t(o\mid s),
\end{equation}
where $N_m$ denotes the number of possible outcomes of observation modality $m$. This hidden-state-specific entropy is normalized by its maximum possible value $\log N_m$, yielding a dimensionless quantity in the interval $[0,1]$, with the two limiting cases corresponding to deterministic and uniformly distributed observation predictions, respectively. The $m$th modality entropy at time step $t$ is finally obtained by averaging the normalized entropy across the $N^m_\text{states}$ hidden states represented in  $A^{m}_t(o\mid s)$, as follows:
\begin{equation}
\mathcal{L}^m_t = \frac{1}{N^m_\text{states}\log N_m} \sum_{s=1}^{N^m_\text{states}}\mathcal{L}^m_t(s).
\end{equation}
The associated temporal evolution is obtained by averaging the corresponding entropy trajectories across the $N_r$ random seeds, as
$\overline{\mathcal{L}^{m}}_t=\frac{1}{N_r}\sum_{n=1}^{N_r}[\mathcal{L}^{m}_t]_n$, with corresponding $\mathrm{SEM}(\mathcal{L}^{m}_t) =\frac{\operatorname{std}_n([\mathcal{L}^{m}_t]_n)}{\sqrt{N_r}}$. The temporal evolution reported in Figure~3 of the main text displays the mean trajectories for the two observation modalities employed in the first case study, together with uncertainty bands extending up to $\pm 2\,\mathrm{SEM}$.

\section{Implementation of Bayesian model reduction}
Bayesian model reduction (BMR) is employed in the first case study to identify and remove redundant hidden-state representations after learning through interaction with the physical system~\cite{friston2016bayesian,friston2017active}. Specifically, BMR is applied to the hidden-state factor associated with the load position, with the objective of obtaining a more parsimonious generative model. Figure~4a of the main text illustrates a representative evolution of the perception--action loop, in which $4$ load-location hypotheses are removed at time step $600$. The corresponding observation model for the signal-norm modality before and after reduction is shown in Figure~4d.

The procedure follows the post-hoc model-reduction framework introduced in~\cite{smith2020active}. In this approach, a full generative model is first learned from observations, after which alternative reduced models are evaluated analytically by modifying the prior concentration parameters while retaining the posterior concentration parameters acquired during learning. This avoids retraining the agent for every candidate model and allows model complexity to be assessed directly from the learned posterior.

In the present implementation, the relative evidence of each candidate model is evaluated from the posterior Dirichlet concentration parameters accumulated from state occupancies. Specifically, the Bayesian learning procedure discussed in Section ``Method'' of the main article is applied to the parameters of $\mathbf{D}$ rather than to those of  $\mathbf{A}$. For notational simplicity, we consider a generative model with a single hidden-state factor and a single observation modality. 

The initial-state prior  $\mathbf{D}\in\mathbb{R}^{\mid\mathcal{S}\mid}$, encoding the initial belief $p(S_0;\mathfrak{D})$, is parametrized by the categorical probability vector \mbox{$\mathfrak{D}\in\mathbb{R}^{\mid\mathcal{S}\mid}$}, such that:
\begin{equation}
S_0;\mathfrak{D}\sim\text{Cat}(\mathfrak{D}),\quad \mathfrak{D}\sim\text{Dir}(\mathfrak{d}),
\end{equation}
where $\mathfrak{d}\in\mathbb{R}^{|\mathcal{S}|}$ collects the positive concentration parameters defining the Dirichlet prior over $\mathfrak{D}$. After Bayesian learning, the variational posterior over $\mathfrak{D}$ is parameterized by $\widehat{\mathfrak{d}}$, as follows:
\begin{equation}
Q(\mathfrak{D})=\text{Dir}(\widehat{\mathfrak{d}}),
\end{equation}
where the fixed-point update rule for $\widehat{\mathfrak{d}}$ reads:
\begin{equation}
\widehat{\mathfrak{d}}^*=\mathfrak{d}+\eta Q^*(S_{t_c}),
\end{equation}
where $Q^*(S_{t_c})$ is the posterior belief over the hidden state at the current-interaction time $t_c$, and $\eta$ is the learning rate. In the implementation used for the BMR simulations, the learning rate associated with $\mathbf{D}$ is set to $\eta=1$.

After learning, candidate reduced models are constructed by modifying the prior concentration vector $\mathfrak{d}$ to impose a small concentration parameter on the hidden states considered for removal. Specifically, let $\mathcal{R}\subseteq\mathcal{S}$ denote the subset of states retained by a candidate model, and let $\mathfrak{d}^{\mathcal{R}}$ denote its corresponding prior concentration vector, defined component-wise as follows:
\begin{equation}
\mathfrak{d}^{\mathcal{R}}_s=
\begin{cases}\epsilon &\text{if } s\notin\mathcal{R},\\
\mathfrak{d}_s &\text{if }  s\in\mathcal{R},\end{cases}
\end{equation}
where $\epsilon>0$ is a small concentration parameter. For the BMR comparison, the prior concentration parameters of the full model are set to $\mathfrak{d}_s=1$ for all hidden states, corresponding to an unbiased prior, while $\epsilon=10^{-3}$ is used for states removal.

Following a standard BMR formulation, the relative change in variational free energy between the candidate reduced model and the original full model can be evaluated analytically~\cite{friston2018bayesian}, in terms of multivariate beta functions:
\begin{equation}
\Delta F_{\mathcal{R}}=\log \beta(\widehat{\mathfrak{d}})+\log \beta(\mathfrak{d}^{\mathcal{R}})-\log \beta(\mathfrak{d})-\log \beta(\widehat{\mathfrak{d}}+\mathfrak{d}^{\mathcal{R}}-\mathfrak{d}),
\end{equation}
where the multivariate beta function $\beta(\,\cdot\,)$ is evaluated as follows:
\begin{equation}
\log \beta(\mathfrak{d})=\sum_{s\in\mathcal{S}}\log\Gamma(\mathfrak{d}_s)-\log\Gamma\left(\sum_{s\in\mathcal{S}}\mathfrak{d}_s\right),
\end{equation}
and $\Gamma(\,\cdot\,)$ denotes the gamma function. This provides an analytical evaluation of the evidence that would have been obtained under alternative priors associated with simpler nested generative models.

For the simulations reported in the main text, the reduced hidden-state factor initially comprises 10 load-position states. Considering $4$ states for removal results in $\binom{10}{4}=210$ candidate reduced models, each corresponding to a different subset $\mathcal{R}$ of $6$ retained states. The selected candidate is the one yielding the smallest $\Delta F_{\mathcal{R}}$. The corresponding $4$ hidden states are then removed from the learned generative model, resulting in a reduced representation of the load-position state space.

The reduction is applied consistently to the different components of the generative model. The selected states are removed from $\mathbf{D}$; their associated columns are removed from the observation model $\mathbf{A}$, eliminating the observation mappings associated with the discarded states; and the corresponding rows and columns of the transition model $\mathbf{B}$ are also removed, together with the action slices associated with the discarded states. The remaining probability distributions are subsequently renormalized. The associated Dirichlet concentration parameters are also pruned consistently with these dimensional reductions. Because the reduced hidden-state factor is directly controlled by an action variable, the actions corresponding to the removed load positions are also deleted from the action space. Following reduction at time step $600$, the simulation is resumed using the reduced generative model for a further 400 physical--digital interactions.

Note that this implementation differs from the BMR application  in~\cite{smith2020active}, as the selected hidden states are explicitly removed from the generative model, rather than being reset while preserving the original dimensionality of the state space. Our BMR application therefore provides a mechanism for post-learning structural simplification. Online Bayesian learning adapts the parameters of the generative model during interaction, whereas BMR subsequently adapts its structure through a hidden-states removal that yields the most favorable evidence trade-off according to the information accumulated during interaction.

\end{document}